\documentclass[12pt, preprint]{aastex}
\usepackage{geometry} 
\slugcomment{Accepted by AJ}

\shorttitle{H$\alpha$ Variability of P Cygni} 
\shortauthors{Richardson et al.} 

\begin{document}
%%\received{} 
%%\accepted{} 

\title{The H$\alpha$ Variations of the Luminous Blue Variable P Cygni: \\
Discrete Absorption Components and the Short S Doradus Phase}
\author{N. D. Richardson\altaffilmark{1}, 
N. D. Morrison\altaffilmark{2},
D. R. Gies\altaffilmark{1},
N. Markova\altaffilmark{3},
E. N. Hesselbach\altaffilmark{4}, and
J. R. Percy\altaffilmark{5}}

\altaffiltext{1}{Center for High Angular Resolution Astronomy, 
Department of Physics and Astronomy, 
Georgia State University, P. O. Box 4106, Atlanta, GA  30302-4106; 
richardson@chara.gsu.edu, gies@chara.gsu.edu} 
\altaffiltext{2}{Ritter Astrophysical Research Center, Department of Physics and Astronomy, University of Toledo, 2801 W. Bancroft, Toledo, OH 43606; nmorris@utnet.utoledo.edu}
\altaffiltext{3}{Institute of Astronomy with National Astronomical Observatory, Bulgarian Academy of Sciences, P.O. Box 136, 4700 Smoljan, Bulgaria; nmarkova@astro.bas.bg}
\altaffiltext{4}{University of Notre Dame, Department of Physics, 225 Nieuwland Science Hall, Notre Dame, IN 46556; ehesselb@nd.edu}
\altaffiltext{5}{Department of Astronomy and Astrophysics, University of Toronto, Toronto, ON, Canada, M5S~3H4; jpercy@utm.utoronto.ca}

\setcounter{footnote}{5}

\begin{abstract}
P Cygni is a prototype of the Luminous Blue Variables (or S Doradus variables), 
and the star displays photometric and emission line variability on 
a timescale of years (known as the ``short S Doradus phase'' variations). 
Here we present new high resolution H$\alpha$ spectroscopy of P~Cyg that
we combine with earlier spectra and concurrent $V$-band photometry to 
document the emission and continuum flux variations over a 24~y time span.
We show that the emission and continuum fluxes vary in concert on 
timescales of 1.6~y and longer, but differ on shorter timescales. 
The H$\alpha$ profile shape also varies on the photometric timescales, 
and we describe the observed co-variations of the emission peak and 
absorption trough properties.  We argue that the episodes of 
photometric and emission brightening are caused by increases in the 
size of the emission region that are related to variations in wind mass loss 
rate and outflow speed.  We find evidence of blueward accelerating, 
Discrete Absorption Components (DACs) in the absorption trough of the H$\alpha$ profile, 
and these features have slower accelerations and longer durations than those 
observed in other lines.  The DAC strengths also appear to vary on 
the photometric timescales, and we suggest that the propagation of the 
DAC-related wind structures is closely related to changes in the overall 
wind mass loss rate and velocity. 
\end{abstract}


\section{Introduction}
Luminous Blue Variables (LBVs or S Doradus variables) are evolved, massive stars. 
LBVs are characterized by large mass loss rates and variability on multiple timescales. The two ``prototypical" Galactic LBVs are $\eta$ Carinae and P Cygni, and they probably represent different extremes of mass loss rate within the scheme of LBV evolution \citep{IdG99}. One of the defining criteria of the LBVs is the observation of a large scale eruption, when the star brightens by several magnitudes. The quiescent times between these eruptions may last centuries. In addition to such rare, giant eruptions, these stars also display lesser photometric and spectroscopic variations on other timescales (e.g.~ Humphreys \& Davidson 1994). \citet{vg01} defines the S Doradus (SD-) phase to be the moderate, long-term, brightening and fading phases.  There are two types of these phases, short and long, with similar characteristics. The short SD-phase is typically on the order of years ($\lesssim$ 10 years), while the longer SD-phase is on a timescale of decades. These phases are thought to originate from changes in the star's photosphere, and both may have the same physical driving mechanism. These long-term variations are observed to differ from cycle to cycle, both in duration and amplitude \citep{ster97}.  

P Cygni (HD 193237, HR 7763, Nova Cyg 1600) remains one of the most fascinating objects in the sky. It was discovered during its first recorded great eruption in 1600 by Willem Janszoon Blaeu, a Dutch chart-maker and mathematician. During this eruption, the star brightened to about 3rd magnitude for about six years, and then it faded from visibility by 1626. It rose again in 1654 to about the same maximal brightness, where it remained for five years. The star faded after this, and although its long-term variability is poorly documented, the star has been slowly brightening to its current magnitude of about 4.8 \citep{IdG99}. The slow brightening may reflect evolutionary changes (de Groot \& Lamers 1992; Lamers \& de Groot 1992; Langer et al.\ 1994).

Long-term photometric monitoring of P Cygni began in the 1980s when \citet{PW83}, \citet{per88}, and \citet{dG90} embarked on extended observing campaigns. Their observations showed that the variations often occur on three characteristic timescales: a short $\sim$ 17 day variation similar to the $\alpha$ Cygni type variations observed in hot supergiants, a $\sim$ 100 day ``quasi-period" similar to that observed in other LBVs, and a long-term cycle (years) attributed to a short-SD phase (de Groot et al.~2001; Percy et al.~2001). 

According to \citet{IdG99}, comprehensive spectral monitoring of P Cygni was started by Luud (1967) and Markova (1993), among others. The first long-term spectroscopic monitoring campaign of P Cygni was presented in seminal papers by Markova et al.~(2001a,b). They found evidence of co-variability of the H$\alpha$ emission line strength and Johnson $UBV$ photometry, indicating a short-SD phase with a quasi-period of $\sim$7 years, although their observations did not fully cover two cycles. The variations were attributed to inversely correlated changes in effective temperature and radius, maintaining a nearly constant luminosity. A similar cycle time was found by \citet{dG01}, who reported on photometric variations which were consistent with a timescale of 5.5 to 8.5 years.

%replacement text from Doug...
In addition to the large scale variations in emission strength, 
Markova (2000) found that there are at least four other kinds 
of line profile variability in the spectrum of P~Cygni.
The most striking of these is the long documented appearance
of blueward-migrating, absorption sub-features that are called 
Discrete Absorption Components (DACs: Israelian \& de Groot 1999;
Markova 2000).  These are generally observed in low and 
intermediate excitation state lines in the optical 
(Markova 2000) and UV spectrum (Israelian et al.\ 1996).
They are frequently detected in the upper sequence of the H Balmer 
lines (principal quantum number $9\le n \le 15$; Markova 2000), 
but to our knowledge, DACs have not been reported before now 
for the absorption component of H$\alpha$.  
DACs are often (but not always) narrow 
(FWHM $\approx 10 - 15$ km~s$^{-1}$) and may be unresolved
in low dispersion spectra.  The DACs tend to appear over 
a radial velocity range of $-90$ to $-200$ km~s$^{-1}$ with an
acceleration of $-0.1$ to $-0.6$ km~s$^{-1}$~d$^{-1}$.  
A recurrence timescale of $\sim 200$~d is sometimes observed 
(Kolka 1983; Markova 1986a; Israelian et al. 1996; Kolka 1998).  
These accelerations are much slower and the timescales  
are much longer than those associated with DACs in the winds of O-stars 
(Kaper et al.\ 1999).  The DACs in the spectrum of P~Cygni may 
form in outward moving and dense shells (Kolka 1983; Lamers et al.\ 1985;
Markova 1986a; Israelian et al.\ 1996), in spiral-shaped co-rotating
interaction regions (CIRs; Cranmer \& Owocki 1996; Markova 2000), or 
in dense clumps in the wind (L\'{e}pine \& Moffat 2008).

%P Cygni is one of the few stars where decades of spectroscopic observations demonstrate the general presence of discrete absorption components (DACs), which are seen as a broad absorption component that appears at low velocity and then evolves to higher (negative) velocity while becoming more narrow (Markova 2000). \citet{IdG99} note that DACs are commonly found in absorption lines in the optical spectrum transitions from many different elements such as H, He, C, N, and Fe. While these are low contrast features that are intrinsically narrow (FWHM of 10-15 km s$^{-1}$), they are readily identified through high resolution spectroscopy. There have been several proposed origins of the DACs including outward moving dense shells in the wind and slowly evolving perturbations in the wind. Co-rotating Interaction Regions (Cranmer \& Owocki 1996) have been ruled out by an upper limit of the rotation period and the long timescale of the DACs in comparison (\citet{IdG99}; Markova 2000).

In this paper, we present new high resolution H$\alpha$ spectroscopy, which we combined with previous measurements by \citet{mar1a} to explore the characteristics of P Cygni's short SD-phase.  We also compare this with archival Johnson $V$ photometry and new observations obtained by AAVSO observers. Section 2 describes our observations. In Section 3, we present our analysis of long-term variations of the continuum and the H$\alpha$ equivalent width. We describe the H$\alpha$ profile morphology changes and DAC propagations in Section 4. Our discussion and conclusions are presented in Section 5.

\section{Observations}
\label{obs}

We obtained 126 new spectroscopic observations of P Cygni using the Ritter Observatory 1~m telescope and \'echelle spectrograph \citep{mor97} between 1999 June 7 and 2007 October 30. These high resolving power ($R = 26,000$) spectra were reduced by standard techniques with IRAF\footnote{IRAF is distributed by the National Optical Astronomy Observatory, which is operated by the Association of Universities for Research in Astronomy, Inc., under cooperative agreement with the National Science Foundation.}. Observations collected prior to 2007 were taken using the setup described in \citet{mor97}. These observations record a 70 \AA\ range in the order centered on H$\alpha$, and they typically have a signal-to-noise ratio between 50 and 100 per resolution element in the continuum. Observations made during the calendar year 2007 were taken with the same spectrograph, except the camera was a Spectral Instruments 600 Series camera, with a front-illuminated Imager Labs IL-C2004 4100$\times$4096 pixel sensor (15$\times$15 micron pixels). To maintain consistency with the older observations, the camera was operated with 2$\times$2 pixel binning. The newer observations recorded a larger portion of the order centered on H$\alpha$ and typically reached a signal to noise ratio between 50 and 100 per resolution element in the continuum. We trimmed these spectra so that the wavelength range was the same as in the older data. The spectra taken after 2002 September have poor wavelength calibration due to problems with the Th-Ar lamp. In order to use these spectra for kinematical measurements, the telluric H$_{\rm 2}$O lines in the vicinity of H$\alpha$ were fitted to improve the solution. This worked for most cases, but the errors associated with the telluric re-calibration are roughly $\pm 3$ km s$^{-1}$, compared with the errors for earlier data of $\pm 1$ km s$^{-1}$.

We collected $V$-band photometry from three sources. The first was from \citet{mar1a}. This provided concurrent photometry for the H$\alpha$ data previously published. The second set came from \citet{per01}\footnote{Available for download at http://schwab.tsuniv.edu/papers/paspc/pcyg/pcyg.html}. These observations also ended at nearly the same time as the first data set. Finally, we downloaded the photoelectric photometry in the $V$-band from the American Association of Variable Star Observers (AAVSO). The AAVSO data are helpful in understanding the long-term trends, and we only used data where measurements of the check and comparison stars differed from expected values by less than 0.05 mag. The errors in the AAVSO measurements are typically around 0.01 mag, comparable to those of Markova et al.~(2001) and \citet{per01}. The combined set contains 3142 measurements from 1985 to 2009.

\section{The Long-Term Photometric and H$\alpha$ Equivalent Width Variability}

\citet{mar1a} found that the H$\alpha$ line flux, obtained by correcting the observed equivalent widths for the changing continuum, varied in concert with the $V$-band flux over the period from 1989 to 1999 (see their Fig.~3). Here we extend their work by considering the long-term photometric and H$\alpha$ variations through 2007. Figure 1 shows the large time span of available photometry. The light curve over this interval shows rather modest, $\pm 0.1$ mag variations, consistent with the star's classification by van Genderen (2001) as a ``weak-active'' LBV. It exhibits the kind of variability associated with a short SD-phase, similar to that reported by Markova et al.~(2001a).  The short SD-phase is most evident in the data prior to 2000, when the star experienced two fadings of $\approx 0.1$ mag \citep{mar1a, per01}. The amplitude of this long-term variability decreased in subsequent years, which indicates that the properties of the short SD-phase change with time.  We made a fit of the very-long-term trend and found a brightening rate of $\approx 0.17 \pm 0.01$ mag century$^{-1}$ (overplotted in Fig.~1). This rate is consistent with the very-long-term trend of $0.15 \pm 0.02$ mag century$^{-1}$ documented by \citet{deG92}. 

\begin{figure} 
\begin{center} 
\includegraphics[angle=90, height=12cm]{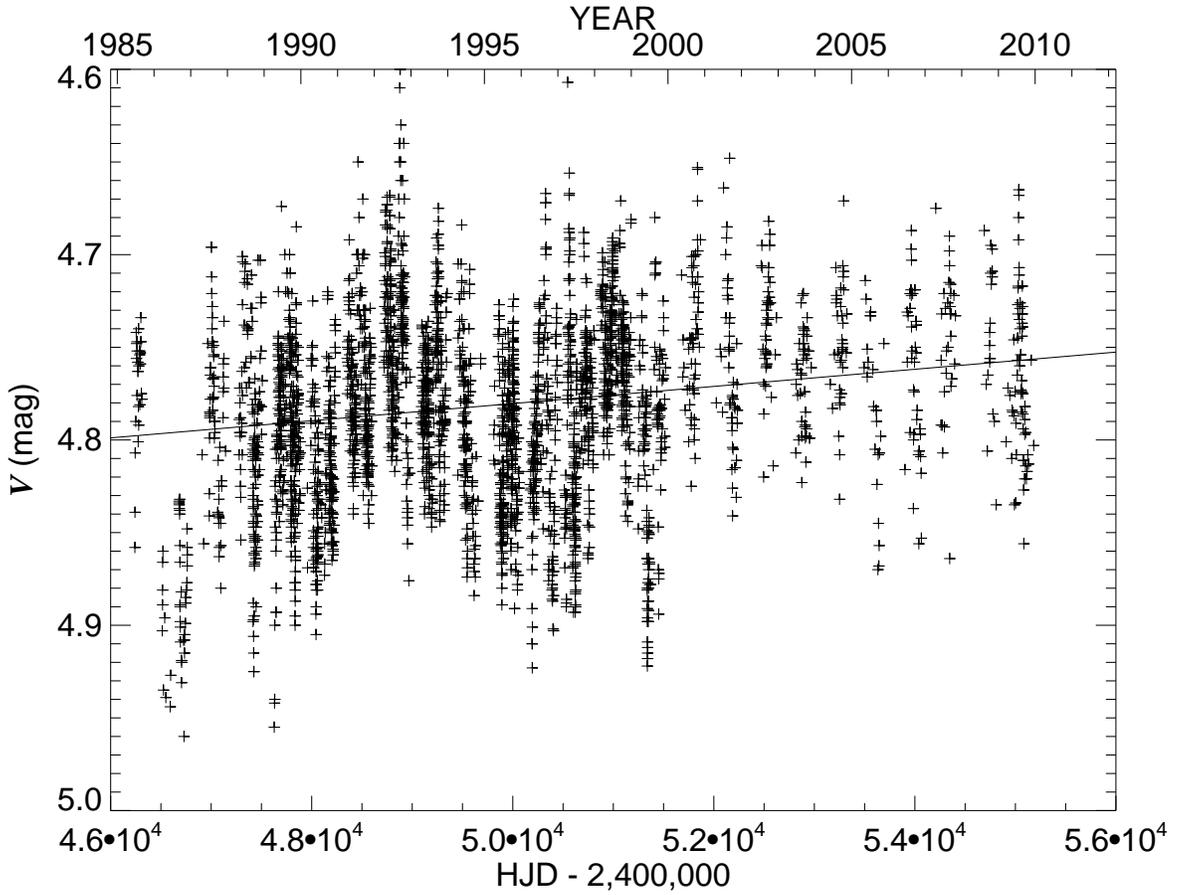}
\end{center} 
\caption{The $V$-band variability of P Cygni between 1985 and 2009. The long-term changes are representative of the short SD-phase. We overplotted the fit for the very-long-term brightening of the star with a solid line.} 
\label{fig1} 
\end{figure}

\placefigure{fig1}

Figure 2 presents the ``dirty" discrete Fourier Transform (Roberts et al.~1987) of the 24.5 y $V$-band photometry with the long-term brightening (Fig.~1) removed. There are no individual significant peaks in the periodogram, but there is a general tendency for more power to appear at the longer timescales (lower frequencies). Thus, the longer timescales of the short SD-phase variability tend to dominate the light curve.

\begin{figure} 
\begin{center} 
\includegraphics[height=15cm]{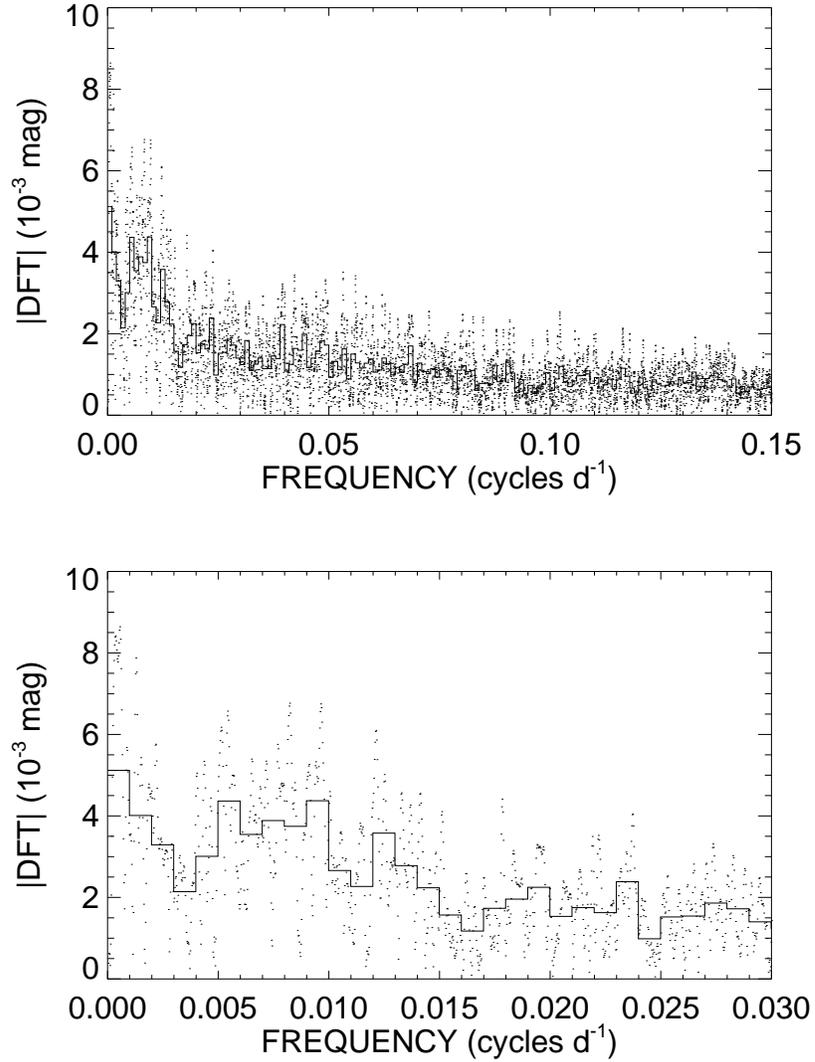}
\end{center} 
\caption{Discrete Fourier transforms of the $V$-band photometry. The bottom panel shows a close up of the long timescale region that is associated with the short-SD phase and 100~d timescales of variability. The overplotted stair steps show the amplitudes rebinned into increments of 0.001 cycles d$^{-1}$.} 
\label{fig2} 
\end{figure}

\placefigure{fig2}

We measured H$\alpha$ emission strength for both the new and originally reported
spectra (Markova et al.\ 2001a) for a total of 158 measurements covering the interval from 1994 to 2007. 
Equivalent widths of the full H$\alpha$ profile (including both the blue absorption and large emission component) were measured in the same manner as done by \citet{mar1a} in order to keep the data sets mutually consistent. The only improvement is that telluric H$_2$O lines in the vicinity of H$\alpha$ were removed by means of a template fitting procedure ({\tt telluric}) in IRAF. This correction resulted in equivalent width increases of less than 2\%. This is much smaller than the typical measurement error of 6\% (as found by comparing equivalent widths from closely spaced observations, $\Delta t < 2$ d, where the variability of this star is minimal). Since the available wavelength range around H alpha does not extend beyond the
electron scattering line wings to the actual continuum levels, a multiplicative constant was used to retrieve the full equivalent width of the line. This correction, $W_{\lambda}$(net) = 1.096 $W_{\lambda}$(Ritter), accounts for unseen line wing flux and unmeasured flux lying below our continuum placement (over an integration range of 6531.5 to 6593.5\AA) and is identical to that adopted by \citet{mar1a} for the Ritter data. The heliocentric Julian dates and net adjusted equivalent widths are tabulated in columns 1 and 2 of Table~1.

The actual line flux of H$\alpha$ can be estimated by correcting the measured equivalent widths for the changing continuum. In order to make this transformation, we averaged the $V$-band measurements made within 20 days of each spectroscopic measurement. This time span was chosen to cancel any small but fast variations and to include enough measurements for a reliable average. We compared all the photometry measurements to a benchmark $V$=4.8 to remain consistent with the flux correction adopted by Markova et al.~(2001a). The equivalent widths were corrected using the relationship 
$$W_{\lambda}({\rm corr})= W_\lambda({\rm net})10^{-0.4(V(t) - 4.8))}.$$ 
The averaged $V$ magnitudes and flux corrected equivalent widths are given in columns 3 and 4 of Table 1. If no $V$ magnitude was available within $\pm 20$ days, then no correction was applied, which affects 18 of our measurements.  These correction factors are usually small ($\approx 4\%$) and comparable to the photometric scatter within each time window. 

We show the temporal variations in the flux corrected equivalent widths in Figure 3. The plot includes earlier measurements from \citet{mar1a}, the new measurements from Table 1, and some additional measurements from 2005 to 2007 from Balan et al.~(2010). There are two maxima (occurring around 1992 and 2002) that are separated by $\approx 10$ years, which is longer than the reported lengths of the short SD-phase found by \citet{mar1a}, de Groot et al.~(2001), or Percy et al.~(2001). Furthermore, the rise and fall around the peak in 1992 are steeper than that for the 2002 peak. There is also ample 
evidence of faster variability within each observing season 
that appears to be unrelated to the longer term trends.

A visual comparison of the $V$-band photometry in Figure~1 with 
the flux-corrected H$\alpha$ equivalent widths in Figure~3 
immediately shows some variations in common.  We found that 
the relative flux (from the time interpolated magnitude) is 
positively correlated with the corrected equivalent width. 
A linear fit yields a slope of 
$\triangle (F/<F>) / \triangle (W_\lambda/<W_\lambda>) = 0.16\pm 0.01$, 
confirming the visual impression of co-variability. 
In order to compare directly the photometry and H$\alpha$ 
equivalent widths, we removed the long-term linear trend from 
the photometry (Fig.~1), performed a running average of the 
photometry differences using a Gaussian weighting scheme 
parametrized by a Gaussian FWHM, transformed the resulting flux 
differences into a variation in \AA ~units according to the 
correlation slope given above, and then added the mean  
equivalent width to the final result.  We made a number of 
trial comparisons by varying the adopted Gaussian FWHM to 
smooth the photometry, and the best fit with FWHM = 598~d is 
shown as a solid line in Figure~3.  For completeness, we also
plot a similar running average of the H$\alpha$ equivalent 
widths as a dashed line.  The agreement between the temporally
smoothed photometric and flux corrected H$\alpha$ variations is 
striking and it appears to confirm the positive correlation 
first noted by Markova et al.\ (2001a).  
The fact that smoothing parameter values smaller than 598 d 
yield worse fits suggests that the co-variations are less correlated 
on shorter timescales.  Taken at face value, this result indicates 
that the continuum and H$\alpha$ emission fluxes sample structures 
in the wind in different ways (probably because of 
different sites of formation in the outflow). 
Finally, we note that the correlation also exists between the 
running averages of the continuum flux and the uncorrected 
equivalent widths $W_\lambda$(net), so the covariations are 
unrelated to the flux correction procedure. 

\begin{figure} 
\begin{center} 
\includegraphics[angle=90,height=12cm]{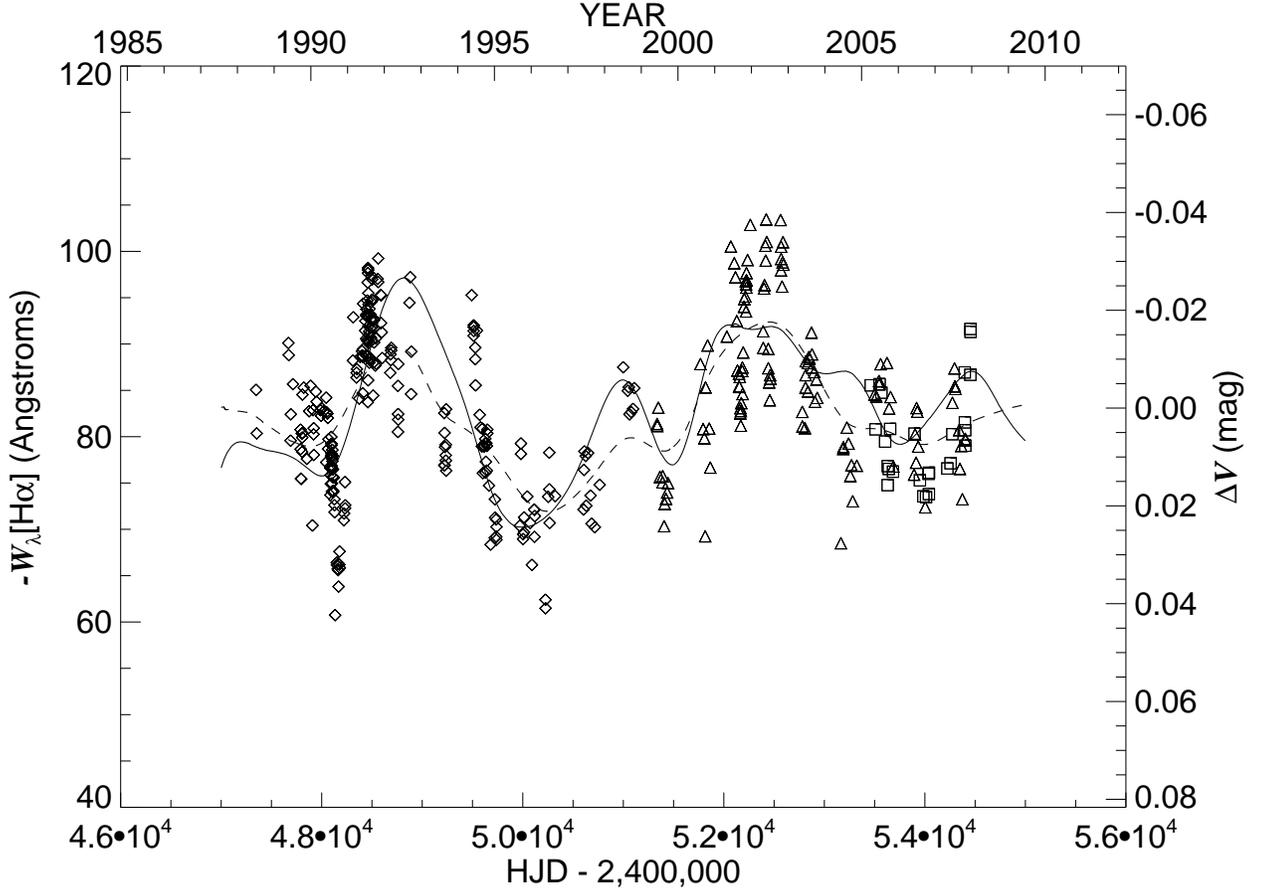}
\end{center} 
\caption{A direct comparison of the smoothed $V$-band photometry (solid line) and the flux corrected H$\alpha$ equivalent widths. The photometric light curve is a running average of the $V$-band flux that was re-scaled and offset to match the equivalent width variations (see text), with the scale of the differential light curve given on the secondary y-axis. A running average of the corrected equivalent width measurements is also overplotted as a dashed line. The equivalent width measurements from Markova et al.~(2001a) are denoted by diamonds, our new measurements are represented as triangles, and the measurements of Balan et al.~(2010) by squares.} 
\label{fig3} 
\end{figure}

\placefigure{fig3}

\section{H$\alpha$ Profile Morphology Variability}

\subsection{Emission Component Changes}

The large H$\alpha$ equivalent width variations described in \S 3 are accompanied by changes in the morphology of the profile. We present two individual spectra in Figure 4 that represent the extrema of the equivalent widths observed (a minimum at HJD 2,450,004, plotted with a dashed-dot line, and a maximum at HJD 2,452,070 plotted with a solid line). It is clear that the profile experiences a change in the peak emission intensity, the line width, the net profile velocity, and the shape of the blue absorption trough. For each of the spectra collected at Ritter Observatory we measured the peak intensity above continuum level, $I_p$, which we corrected for the changing continuum level in the same manner as the equivalent width (\S 3), the FWHM (profile width) of the emission portion of the profile above continuum, and a relative radial velocity $\triangle V_r$, derived by cross-correlating each profile against an unweighted average of all the spectra obtained at Ritter Observatory.  We chose to use a cross-correlation technique because this method is model-free and is most sensitive to the steep emission line wings, resulting in a measure similar to a FWHM bisector velocity. The resulting measurements of FWHM and $\triangle V_r$ are shown for these two profiles in Figure 4 with horizontal and vertical lines, respectively.

\begin{figure}
\begin{center}
\includegraphics[angle=90,height = 12cm]{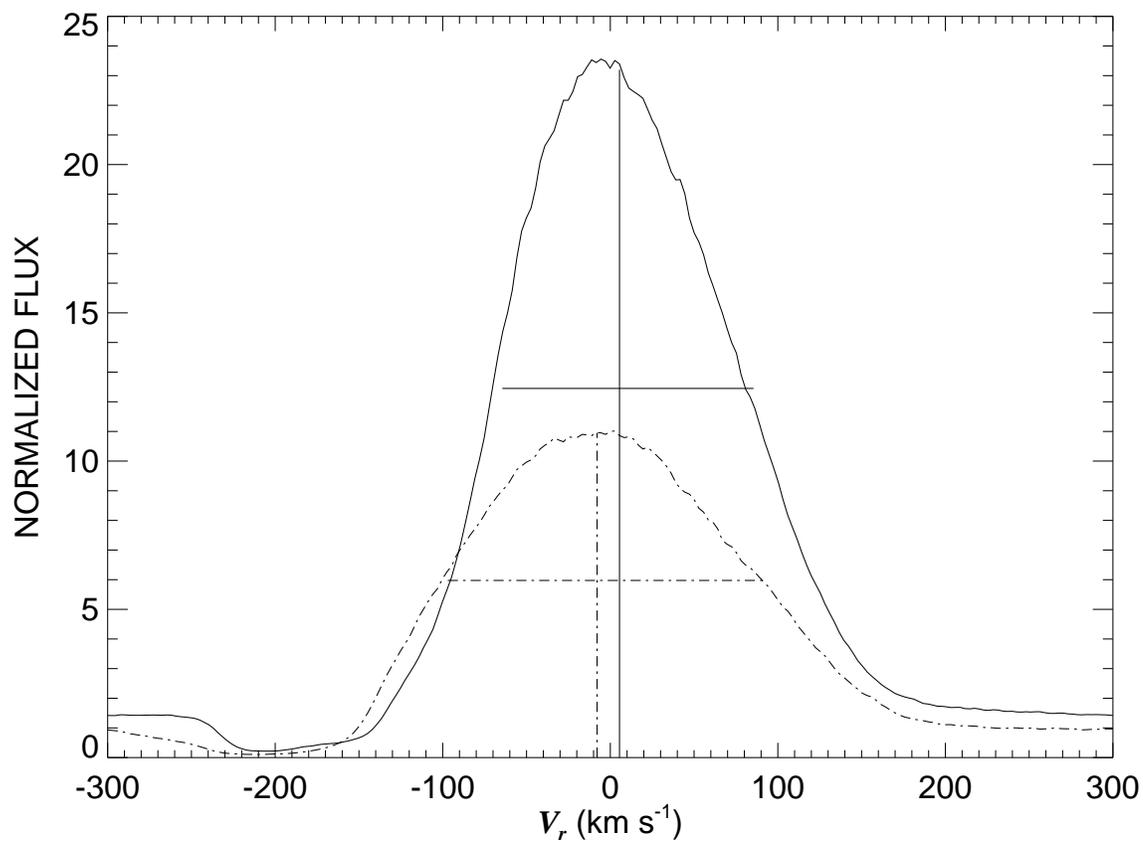}
\end{center}
\caption{A comparison of two observed H$\alpha$ profiles corresponding to extremes of the emission variability (maximum and minimum in corrected equivalent width). The spectrum plotted as a solid line is from HJD 2,452,070, while that shown as a dash-dot line is from HJD 2,450,004. Vertical and horizontal line segments show the velocity offset $\triangle V_r$ and the FWHM range, respectively.}
\label{fig4}
\end{figure}

\placefigure{fig4}

All these measurements are given in columns 5, 6, and 7 of Table 1, and are plotted as a function of time in the three panels of Figure 5. We see that times of strong emission (for example, HJD 2,452,500; see Fig.~3, MJD 5.25$\cdot 10^4$) correspond to profiles with the largest peak intensity and smallest FWHM. We also see a small radial velocity shift that is correlated with the long-term variations. The profile had the largest (most positive) velocity when the line flux at the position of the emission peak was strongest, which was also when the profile showed the smallest FWHM (Fig.~5). This is likely due to changes in the P Cygni absorption component. When the profile has the most emission, the blue absorption portion appears to shift to a more positive velocity and removes more of the blue side of the emission peak (see Fig.~4), and thus, the net radial velocity tends toward a larger (more positive) value at those times. We find that the measurement errors for $\triangle V_r$ and the FWHM are about $1$ km s$^{-1}$, which adds in quadrature to the calibration errors discussed in \S2, to yield net errors of approximately $\pm 1.5$ km s$^{-1}$ for most of the data, and $\pm 3.2$ km s$^{-1}$ for data taken after 2002 September. The errors for $I_p$ are on the order of $3-5\%$. 

Kashi (2010) has suggested P Cygni is a binary system with a fainter B-type companion and that small long-term radial velocity variations due to reflex motion might be observed in extended high resolution spectroscopic observations. This cannot be the explanation for the $\triangle V_r$ changes we observe, since the H$\alpha$ emission is formed over a volume that is much larger  in radius than the predicted semimajor axis of the putative orbit of the P Cygni primary star. Wind gas leaving the star at any instant would have a Keplerian orbital component 
that decreases with distance from the center of mass.
As the gas packet moves out 
to the radius where H$\alpha$ becomes optically thin and emits the photons we observe 
(at $\approx 10 R_\star$ and larger; see below), the radial outflow component will increase 
by radiative driving while the orbital motion component will drop with distance to conserve
angular momentum.  Thus, at the large distance of line formation, the gas motion will
be almost completely radial. 
If the putative companion is to be found from radial velocity variations of this star, then detailed analyses of photospheric or wind lines formed very close to the star will need to be analyzed. Further, these radial velocity variations are not strictly periodic, and cannot be considered orbital motion. Lastly, given the method of measuring these velocities, the measured radial velocity is at least partially due to morphological changes in the H$\alpha$ line profile.

\begin{figure} 
\begin{center} 
\includegraphics[height = 15cm]{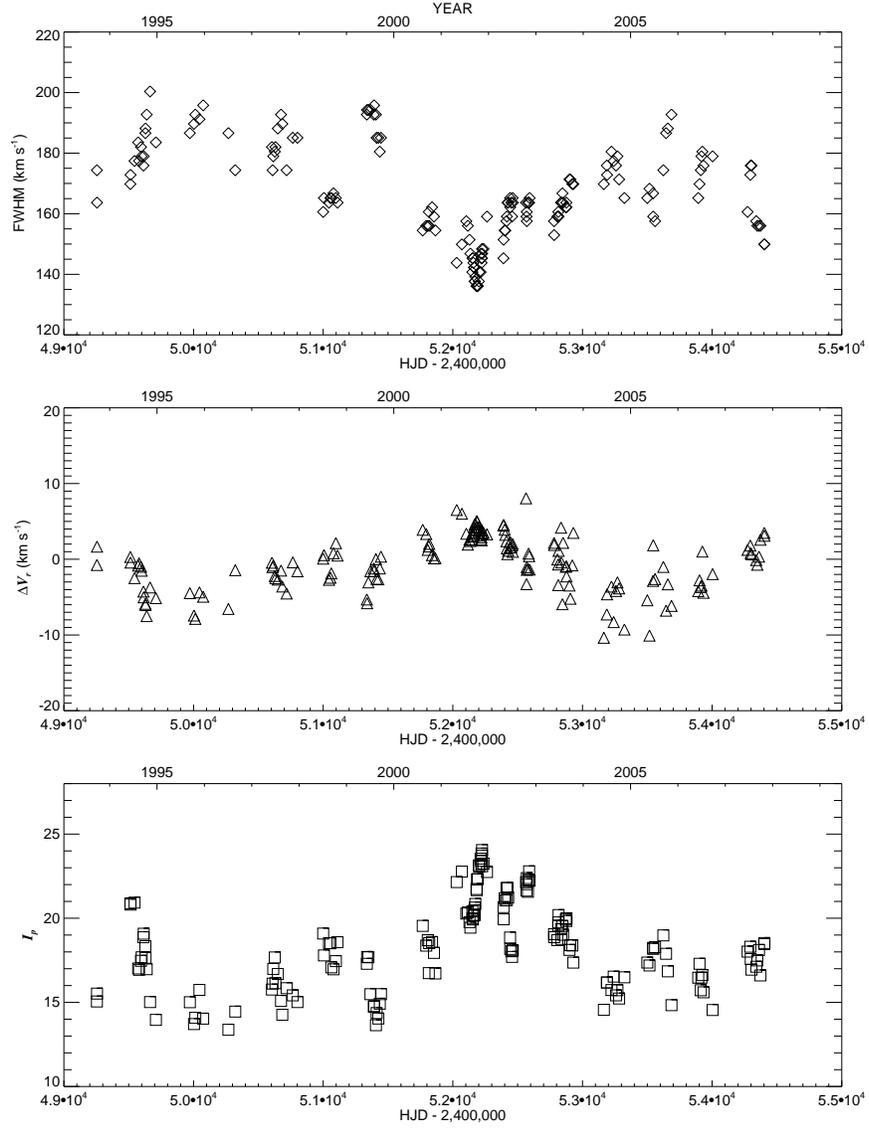}
\end{center} 
\caption{The long-term variability of the FWHM (top panel), the relative radial velocity $\triangle V_r$ (middle panel), and flux corrected peak intensity $I_p$ (bottom panel) of the emission peak of the H$\alpha$ profile. The long timescales of variability are similar to those seen in the $V$-band photometry and the H$\alpha$ equivalent widths (Fig.~3).} 
\label{fig5} 
\end{figure}

\placefigure{fig5}

\subsection{Blue Absorption Changes}

We used all of the H$\alpha$ spectra from Ritter Observatory, including those that were measured by \citet{mar1a}, to investigate the variations in the blue absorption trough of the P Cygni profile. This portion of the profile is especially interesting as it is formed in the outflowing gas along our line of sight to the star. In order to emphasize the relative changes in line absorption, we first formed a reference, average high-intensity, minimum-absorption spectrum, as follows. At each wavelength step, we ordered the time-series by intensity and then constructed the average of all the intensities falling between the 90th and 95th percentiles at that wavelength step. This removed any spurious peaks caused by cosmic rays from contaminating the minimum absorption average. 
We then divided each of the spectra by this reference spectrum to form a matrix of quotient spectra.
Since we are interested in the variability of the central absorption, and not that of the far wings, and because the line wings never reach the continuum in the region we recorded, the quotient spectra had a depressed continuum. We then re-normalized these 
to a unit continuum (outside of the velocity region $\pm 500$ km~s$^{-1}$). These spectra are illustrated in a gray-scale dynamical spectrum in Figure 6. In this figure, we present each quotient spectrum as a function of radial velocity and time with a gray-scale intensity between the minimum value (black; $0.14$ in the quotient) and maximum value (white; 1.75 in the quotient) based upon a linear time interpolation between the nearest observations (indicated by arrows). The low absorption reference spectrum is displayed for comparison in a panel below the gray-scale image. For simplicity, these quotient spectra were not corrected for the variable continuum flux since we are interested in both emission and absorption changes.

\begin{figure}
\begin{center}
\includegraphics[height = 15cm]{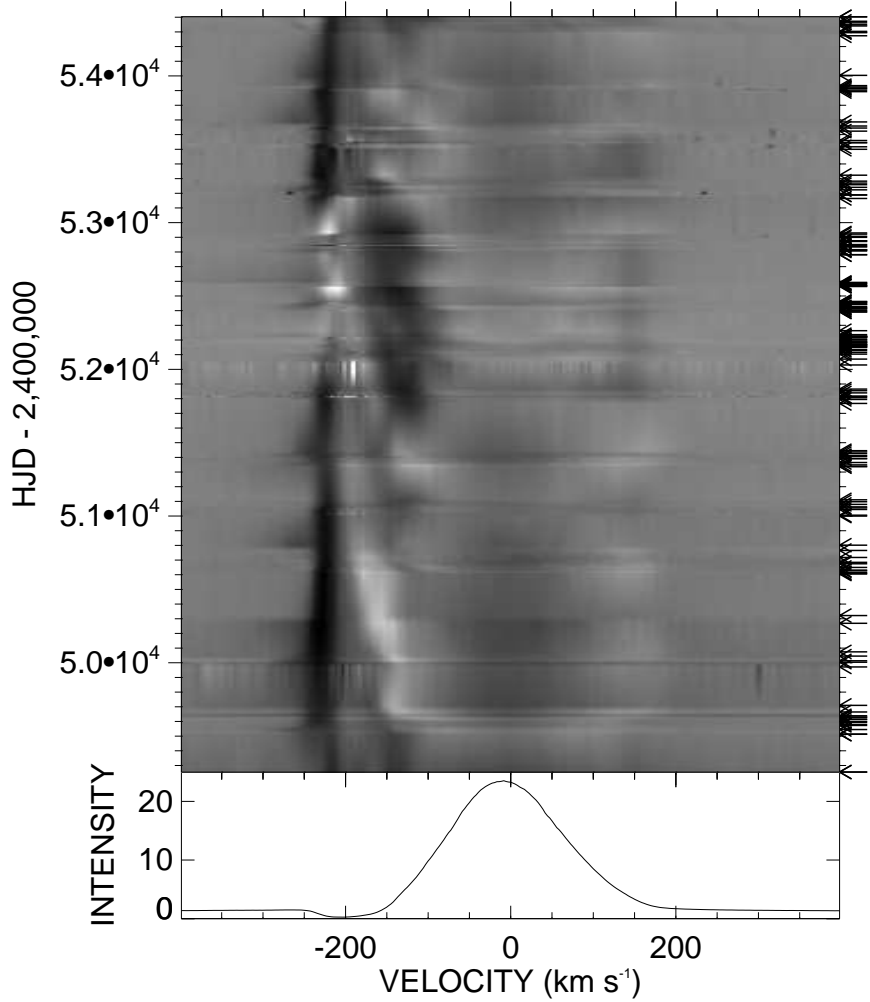}
\end{center}
\caption{A dynamical spectrum showing the quotient of the H$\alpha$ spectra and a constructed low absorption profile (shown in the lower panel) as a function of radial velocity and time. The blueward moving features are the DACs, which are almost always present in H$\alpha$. The gray-scale image is formed by a time interpolation (over 100~d in some rare cases) to fill in observational gaps in the data. Arrows on the right of the diagram indicate the dates of observation. The range of the plot is from $0.14$ (black) to $1.75$ (white) in intensity ratio. The original spectra, as well as the maximum intensity profile, are available in a supplementary file in the online version.}
\label{fig6}
\end{figure}

\placefigure{fig6}

We need to bear in mind that the low absorption spectrum was formed by different
subsamples at each wavelength point, and this has important consequences for the 
appearance of the dynamical spectrum.  For example, inspection of Figure~4 shows that 
blue absorption core can extend to high negative velocities (dash-dot line) while at other times
the blue absorption is limited to moderate velocities (solid line).  Thus, the construction
of the low absorption spectrum will be dominated by the latter examples in the 
vicinity of the blue absorption edge, and in our collection of H$\alpha$ spectra, 
the more extended blue absorption occurred much more frequently.  Consequently, 
the quotient spectra in Figure~6 appear to be dominated by a blue absorption feature, 
near $-220$ km s$^{-1}$, except near HJD~2,452,500 (MJD 5.25$\cdot 10^4$) when the blue edge moved to a more 
positive velocity.  This feature is due entirely to our selection of spectra from 
around HJD~2,452,500 in making the low absorption spectrum. 

We also see evidence in Figure~6 of several blueward moving, absorption features 
(primarily between $-100$ and $-200$ km s$^{-1}$) that appear morphologically similar 
to the Discrete Absorption Components (DACs) observed previously in other spectral 
lines (Israelian \& de Groot 1999; Markova 1986a, 2000).  Figure~7 is a montage of a subset of the
quotient spectra. It shows how the DAC (center) moved progressively 
blueward over this time span ($\approx 800$~d).  There are times where the regions 
between successive DACs appear bright in the dynamical spectrum, which correspond 
to those cases (usually sparsely sampled in time) where the flux was higher than the 
mean in the 90\% to 95\% part of the flux distribution that defined the minimum 
absorption spectrum.  Finally, we see in the gray-scale image of Figure 6 the 
long-term variations in peak intensity ($I_p$) and wing extension (FWHM) that are 
associated with the short SD-phase, equivalent width variations (Fig.~5).

\begin{figure}
\begin{center}
\includegraphics[height = 12cm]{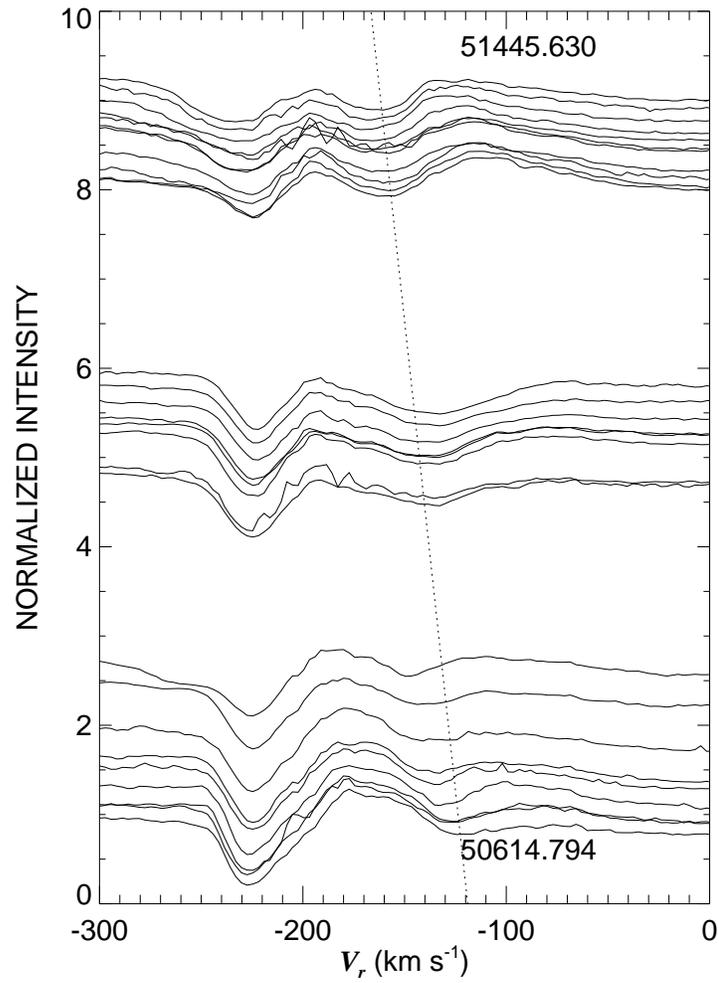}
\end{center}
\caption{A series of quotient spectra offset according to the time of observation (HJD $-$ 2,400,000 indicated for the first and last spectrum in this sequence). The DAC present migrated from approximately $-125$ km~s$^{-1}$ to $-160$ km~s$^{-1}$ over this interval. The dotted line represents a linear fit to the measured centroid velocities for this DAC.}
\label{fig7}
\end{figure}

\placefigure{fig7}

We measured the variability of the quotient spectra by calculating the 
standard deviation at each pixel of velocity space. This standard deviation 
spectrum is shown in Figure~8.  There is a broad feature centered at rest 
velocity which is associated with the varying peak height of the profile. 
Another feature is present at $\sim +150$ km~s$^{-1}$, which could be caused 
by either the variations in the profile width (FWHM) or red emission wing 
variability from traveling bumps (Markova 2000). 
The largest two features are from the DACs (seen as a broad peak around 
$-140$ km s$^{-1}$) and from the variations present near the blue edge
of the absorption core (visible as a peak centered at $-220$ km s$^{-1}$). 

\begin{figure}
\begin{center}
\includegraphics[angle=90,height = 12cm]{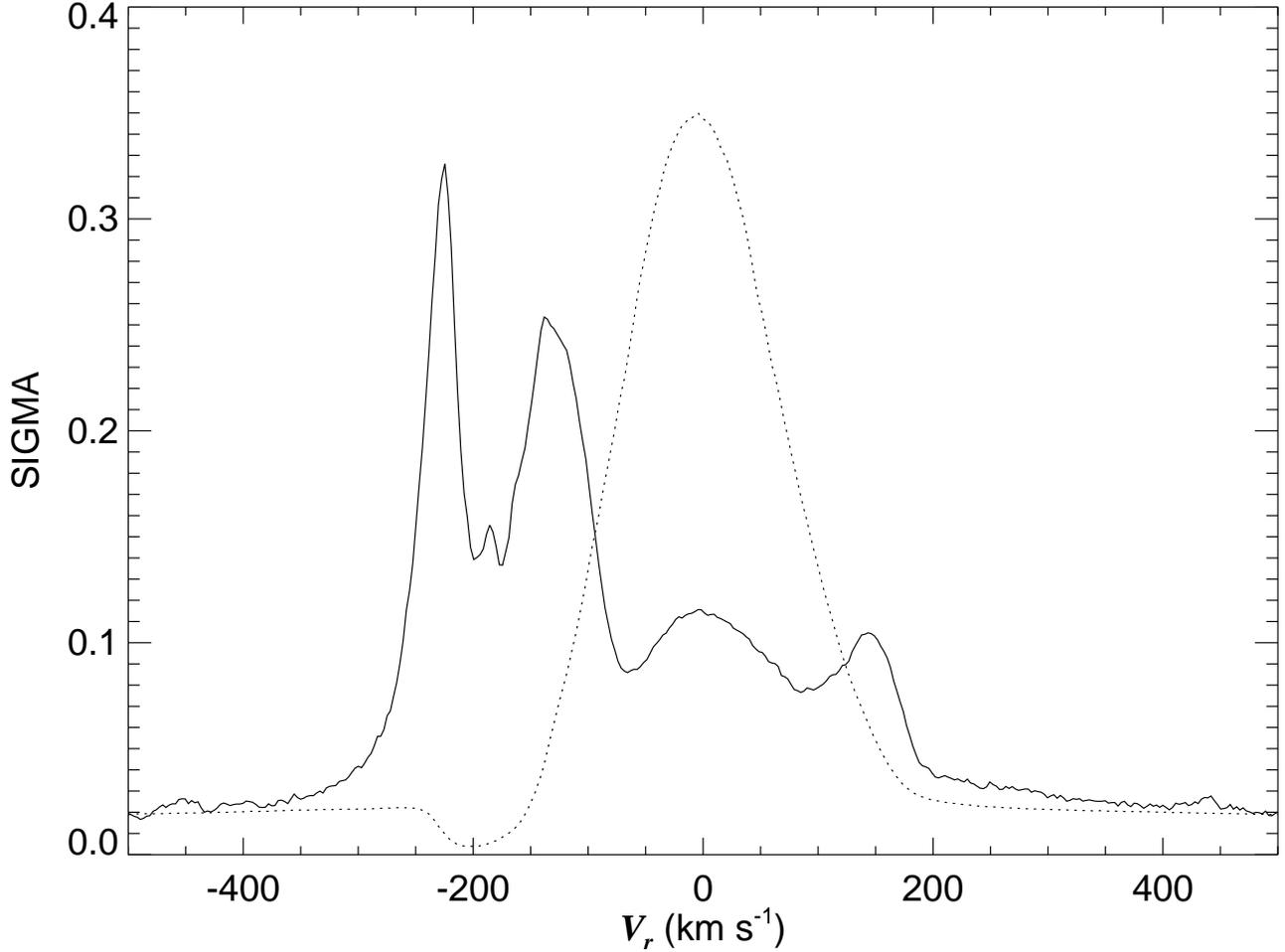}
\end{center}
\caption{The pixel-by-pixel standard deviation of the quotient spectra (solid line). We also show the low absorption reference profile (dotted line; re-scaled to this range) to highlight those parts of the profile where the largest relative variations are occurring: near the terminal velocity blue edge (near $-220$ km s $^{-1}$), over the range traversed by the DACs (centered near $\approx -150$ km s $^{-1}$), near the emission peak ($0$ km s $^{-1}$), and in the emission wings ($\pm 150$ km s $^{-1}$). The standard deviation of the quotient is larger in the absorption core because the low absorption profile is close to zero there.}
\label{fig8}
\end{figure}

\placefigure{fig8}

Figures 4, 6, 7, and 8 all demonstrate that there are significant changes
observed in the profile near the blue edge of the absorption core. 
The blue-edge velocity of a P~Cygni profile is generally set by the 
position where the absorption core rises to intersect with the 
local continuum, and this velocity corresponds to the fastest moving 
gas projected against the disk of the star.  However, in the 
case of H$\alpha$ in the spectrum of P~Cygni, this location is often 
poorly constrained because absorption may extend blueward with a 
shallow slope (see Fig.~4).  Thus, we decided instead to document 
the kinematical changes near the blue edge by measuring the position 
of the absorption core flux minimum, $V_{r}(\rm min)$, which is 
normally found near $-210$ km~s$^{-1}$ where the slope of the profile
changes sign abruptly.   We determined this position by finding the zero 
crossing in the numerical derivative of a smoothed version of the spectrum.
The S/N ratio was sufficient in all our spectra that the zero of the 
derivation was always well-defined.   This estimate of the minimum flux 
velocity $V_r$(min) is given in column 8 of Table 1, and the errors in 
$V_{r}(\rm min)$ are comparable to the errors associated with the emission 
line kinematic measurements (\S4.1).  This velocity is probably related to 
the wind speed at a location in the wind where H$\alpha$ ceases to be
optically thick.  

It is difficult to measure the radial velocities of the DACs because their morphologies vary and because the absorption may consist of multiple components. We decided to measure a centroid for the DACs wherever possible by means of the relationship
$$V_r{\rm (DAC)}={{\int_{v_1}^{v_2}{v_r (1-Q(v_r)) ~dv_r}}\over{\int_{v_1}^{v_2}{(1-Q(v_r))~dv_r}}} $$
where $Q$ represents the quotient spectrum and $v_r$ is the radial velocity. We adopted a velocity range of $v_1 = -200$ km~s$^{-1}$ and $v_2 = -100$ km~s$^{-1}$ based upon the strongest regions of the standard deviation spectrum plotted in Figure 8. Typical errors in these measurements (from the scatter in densely sampled regions of the time series) are 
$\pm 3$ km~s$^{-1}$. This approach worked for most of the spectra, but some low contrast features were not measured correctly, and are omitted from Table 1 and our analysis. During some epochs, there were multiple DACs present, so 
$V_r{\rm (DAC)}$ represents a weighted average of multiple components. The DAC radial velocities $V_r{\rm (DAC)}$ 
are given in column~9 of Table~1.  Column~10 lists a relative equivalent width for the DAC measured by a 
direct numerical integration of $Q(v_r)$ between $v_1$ and $v_2$. The typical errors associated with these equivalent width measurements are on the order of 5\%, which is similar to the errors associated with the equivalent widths of the profile (\S3). 

We show the temporal evolution of the DAC and blue-minimum flux velocities in Figure~9. 
At some epochs, a long progression of DAC velocities is present. The most well defined sequence began near HJD 2,450,700, and lasted about 800 days (Fig.~7).  This timescale is much longer than that of a typical DAC observed in an O star, where the progression is measured in hours or days. As this particular DAC was relatively narrow and recorded in many spectra over its duration, it is an optimal feature to measure the H$\alpha$ DAC acceleration. From a simple linear fit, shown overplotted in Figures 7 and 9, we measured the acceleration to be $-0.047 \pm 0.002$ km~s$^{-1}$~d$^{-1}$. For comparison, the observed DACs in the spectrum of a normal hot supergiant of similar spectral type, $\epsilon$ Ori (HD 37128; B0~Ia), have an acceleration of $-500$ km~s$^{-1}$~d$^{-1}$ (Prinja et al. 2002). 

\begin{figure}
\begin{center}
\includegraphics[angle=90,height = 12cm]{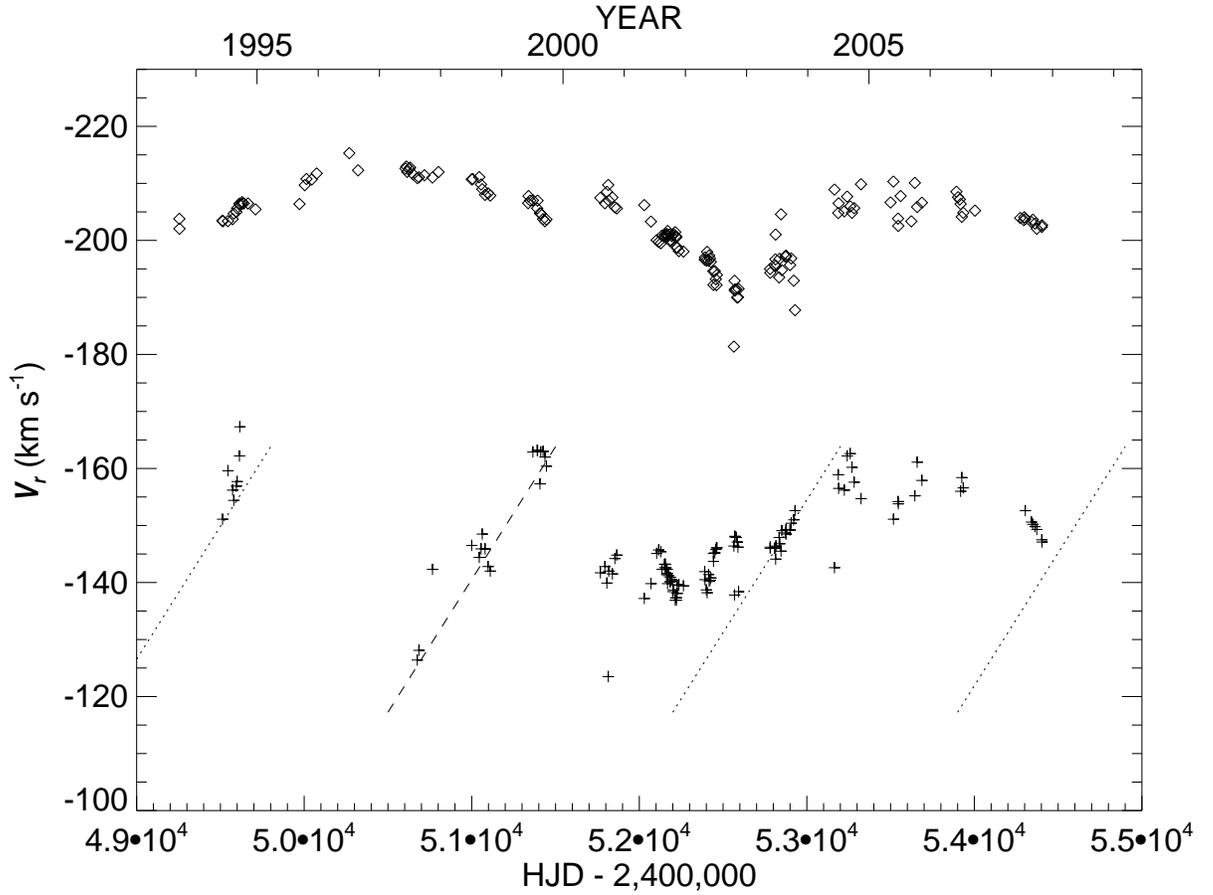}%%DAC_vels1.ps to not include 1700 d relaunch
\end{center}
\caption{The temporal variations of the measured velocities of the DACs, $V_r$(DAC) (plus signs), and the minimum flux wind velocities, $V_r$(min)  (diamonds). The dashed line represents the acceleration fit to the DAC progression near HJD 2,451,000, and the dotted lines represent the same fit translated by intervals of 1700 days, which we derived as a possible recurrence time.}
\label{fig9}
\end{figure}

\placefigure{fig9}

We used the Lomb-Scargle periodogram method (Scargle 1982) to search for a recurrence time in the appearance of the DACs, and we derived a cycle time of 1700~d. This recurrence timescale is shown in Figure 9 where we overplot the linear acceleration of the DAC shown in Figure 7 for three additional epochs over the time span of the Ritter data. There is some evidence that a DAC progression is seen at each of these four epochs. The major deviations from the expected velocity trends occur when multiple components are present. For example, there were two DACs present between HJD 2,451,700 and 2,452,400, and the velocity centroid we measured represents a blend of these components. 
Neither of these two DACs occurred at the expected recurrence time in the 1700~d cycle. 

Our work represents the first detection of DACs in the blue
absorption trough of H$\alpha$, and their properties differ from those 
observed in other spectral lines (Israelian et al.\ 1996; Markova 1986a, 2000).
For example, the recurrence timescale of 1700~d is much 
larger than the 200~d interval found in earlier work, and 
the acceleration we measure is about a factor of 10 smaller 
than measured by others.  We suspect that these differences
are due to the large optical depth of the H$\alpha$ line compared
to that of other lines where DACs have been investigated.  This will mean 
that the radius of optical depth unity is larger for H$\alpha$
(Najarro et al.\ 1997), and consequently any wind structures 
formed at smaller radii will have no affect on the H$\alpha$ 
line formation.  We suspect that observations of other, 
less optically thick lines are more sensitive to the 
detection of DACs formed at smaller radii in the wind of 
P~Cygni where more and faster accelerating structures may exist.

\section{Discussion}

The $V$-band and H$\alpha$ variations we observe need to be
interpreted in the context of current models for the star and its
wind.  Langer et al.\ (1994) discuss the atmospheric properties of
P~Cyg, and they argue that the star is in the LBV phase
where the temperature and helium abundance are increasing,
and the mass and luminosity are decreasing, as the star
evolves towards the Wolf-Rayet phase.  Langer et al.\
emphasize the earlier conclusion from Pauldrach \& Puls (1990)
that the atmospheric parameters are close to a bi-stability
point where the mass loss rate can change by an order of
magnitude with small changes in radius and/or luminosity,
which may explain the great eruptions observed in prior
centuries.  The atmospheric parameters are well established
through a detailed quantitative spectroscopic analysis by Najarro et al.\ (1997)
and Najarro (2001), who find that He is overabundant and
that the mass loss rate is high ($\approx 2\times 10^{-5}$
$M_\odot$~y$^{-1}$ including wind clumping effects) and
wind terminal velocity is low ($v_\infty = 185$ km~s$^{-1}$).  
Najarro et al.\ (1997) derive a systemic velocity of $\gamma = -29$ km~s$^{-1}$, 
and thus our minimum measurement of $V_{\rm min} = -215$ km~s$^{-1}$ 
is consistent with their estimate of $\gamma - v_{\infty} = -214$ km~s$^{-1}$ as this velocity measurement is related to $v_{\infty}$.
They estimate that the continuum forming radius is $76 R_\odot$,
which for a distance of 1.8~kpc implies an angular diameter
of $\theta = 0.39$ mas.  On the other hand, Najarro et al.\ (1997)
predict that the emitting size of H$\alpha$ will be much larger
because of its greater optical depth.  For example, their models
show that there is a local maximum in the wind temperature distribution
(presumably where the recombination processes that form H$\alpha$ peak)
near $r/R_\star = 11$ (see their Fig.~5b).  The corresponding
angular size for H$\alpha$ of $\approx 4$~mas agrees well
with the range of $3 - 7$~mas from H$\alpha$ interferometry
by Balan et al.\ (2010).  Thus, we need to keep in mind that
the H$\alpha$ variations reflect changes over a much larger spatial scale
in the wind than those observed in the $V$-band flux.

Variations in the H$\alpha$ emission equivalent width are related
to changes in both the mass loss rate and the wind velocity.
In a very simplified approach, we can assume that most of the H$\alpha$
flux originates in the optically thick region projected on the sky,
$$f = 2 \pi r_\tau ^2 F(T)$$
were $r_\tau$ is the boundary separating the optically thick and
thin regimes, $F(T)$ is the monochromatic surface flux,
and $T$ is the wind temperature at $r_\tau$ (Najarro et al.\ 1997).
If we assume that the wind is approximately isothermal at this physical
location (a reasonable choice: see Fig.~5b in Najarro et al.\ 1997),
then the emission flux variations are due to changes in the projected
size of the optically thick region,
$$\triangle f / f = 2 \triangle r_\tau / r_\tau.$$
Thus, we expect that the relative variations in angular size
will be only half as large as the emission equivalent width variations,
which is probably consistent with the lack of measurable size
changes in the H$\alpha$ interferometric measurements (Balan et al.\ 2010).

The H$\alpha$ optical depth is dependent on the electron density squared
since the emission is a recombination process.  Thus, we expect that
the optical depth unity boundary $r_\tau$ will always be defined by
the location in the wind with a specific characteristic density, $\rho_\tau$.
We assume that $\rho_\tau$ has an approximately constant value so that 
the effective $r_\tau$ boundary will vary as fluctuations in the wind 
mass loss rate and velocity define the radius where the density reaches $\rho_\tau$.
According to the mass continuity equation, $r_\tau$ is related to this density by
$$r_\tau ^2 = {\dot{M}\over {4 \pi \rho_\tau v}}$$
where $\dot{M}$ is the mass loss rate and $v$ is the wind velocity at the radial distance $r_\tau$.
We can differentiate the mass continuity equation to express the 
radius variation in terms of the changes in $\dot{M}$ and $v$,
$$2 r_\tau \triangle r_\tau =  {1\over {4 \pi \rho_\tau}} \triangle [{\dot{M}/ v}],$$
which we divide by $r_\tau ^2$ to obtain
$$2 {\triangle r_\tau \over r_\tau} = {{\triangle [{\dot{M}/ v}]} \over {[{\dot{M}/ v}]}}.$$
Since we argued above that the flux also varies as $r_\tau ^2$, we can 
then use the relation above to re-write the fractional flux variation in 
terms of logarithmic changes in $\dot{M}$ and $v$, 
$$\triangle \ln f =  \triangle \ln \dot{M} - \triangle \ln v .$$
Since we have observational data on the variations in emission strength
and wind velocity, we can rearrange this equation to solve for the
mass loss variations as a function of flux variations,
$$\triangle \ln \dot{M} / \triangle \ln f = 1 + \triangle \ln v /\triangle \ln f .$$
Puls et al.\ (1996) present a much more detailed analysis of the 
dependence of the emission equivalent width on the wind parameters 
of hot, massive stars.  However, in the limit of high optical depth, 
their expression for the emission flux (their eq.~41) leads to a similar relation, 
$$\triangle \ln \dot{M} / \triangle \ln f = {3\over4} (1 + \triangle \ln v /\triangle \ln f) .$$

We found in the previous section that the H$\alpha$ equivalent width
appears to vary inversely with two quantities related to wind dynamics,
the H$\alpha$ emission peak FWHM and the blue minimum flux velocity $V_r$(min).
Figure 10 quantifies this relationship. 
The upper panel shows the inverse correlation between 
the emission peak FWHM and flux corrected H$\alpha$ equivalent width, 
and if we take FWHM as a proxy for the wind speed, then a linear
fit of natural logarithms of these measures gives 
$\triangle \ln v / \triangle \ln f = -0.66 \pm 0.12$.
We caution that the FWHM is also influenced by the absorption 
component of H$\alpha$, and we showed above (Fig.~4) that the 
absorption component moves inward towards the line core when 
the emission is strong.  Consequently, the apparent decrease in 
FWHM as the emission increases probably results both from a
wind speed decrease and a blue wing decline due to blending with 
the absorption component.   The minimum flux velocity is perhaps 
a more direct measurement of wind speed (at least in our line of sight),
and we show in the lower panel of Figure~10 the co-variations of 
the difference between $V_r$(min) and the systemic velocity of P~Cyg, 
$\gamma = -29$ km~s$^{-1}$ (Najarro et al.\ 1997), 
as a function of the corrected equivalent width.  A fit of the 
logarithmic slope here gives a smaller estimate of 
$\triangle \ln v / \triangle \ln f = -0.22 \pm 0.04$. 

\begin{figure}
\begin{center}
\includegraphics[angle=0,height = 12cm]{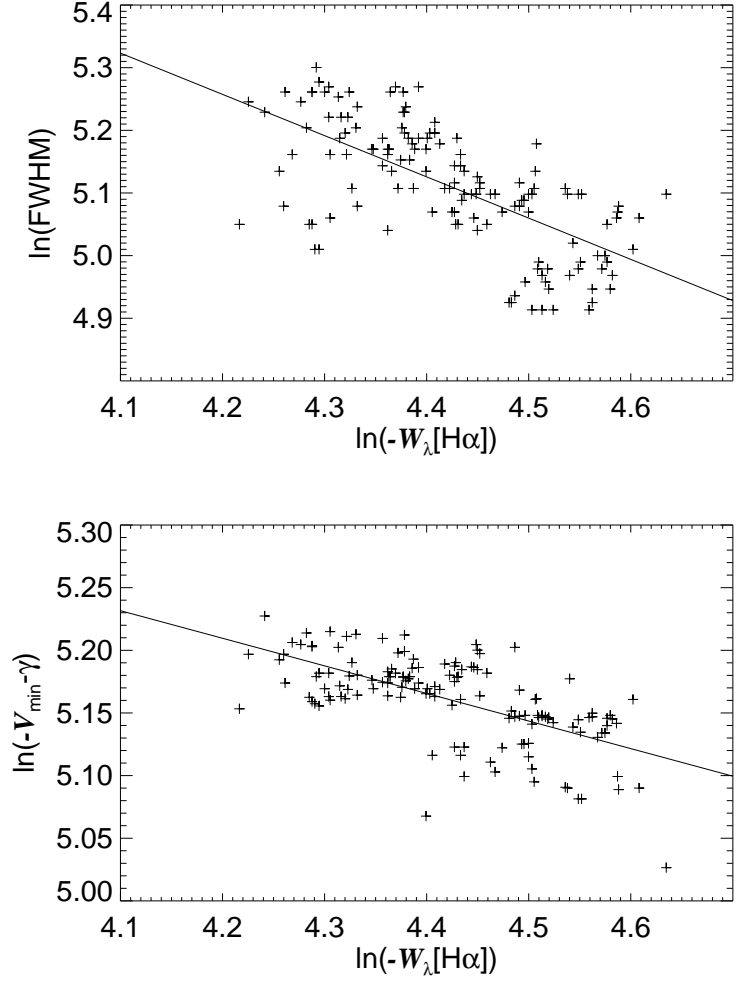}
\end{center}
\caption{A comparison between the flux corrected equivalent width of H$\alpha$ and the FWHM of the emission peak (top panel). The best fit (solid line) has a slope of $-0.66$. The lower panel shows a comparison between the corrected equivalent width of H$\alpha$ and the difference between the minimum flux velocity $V_r$(min) and the systemic 
velocity $\gamma$ of P~Cyg.  A linear fit yields a slope of $-0.22$ (solid line).}
\label{fig10}
\end{figure}

\placefigure{fig10}

If we adopt the minimum flux co-variation result as representative 
of the wind velocity component of variability, then the mass loss rate 
variation we derive from the relation above has an emission flux dependence of 
$\triangle \ln \dot{M} \approx 0.78 \triangle \ln f$.  
Omitting the bottom and top 10\% of the distribution of $W_\lambda$(corr), the derived range in emission strength in our observations of $\pm 14 \%$ probably implies mass loss rate changes of $\pm 11\%$. 
Markova et al.\ (2001a) used the optically thick relation from Puls et al.\ (1996)
to arrive at an estimate of $\pm 9\%$ for the mass-loss variation amplitude. 
We showed above that the relation from Puls et al.\ carries a factor of $3/4$ that is
missing from our simple analysis, and if we use the Puls et al.\ relation instead, 
we arrive at a mass loss rate variation of $\pm 8\%$, confirming the estimate from
Markova et al.\ (2001a).  In this scenario, the H$\alpha$ emission variations result from 
changes in the effective emission radius $r_\tau$ caused by 
variations in the mass loss rate and wind velocity. 
During episodes when the mass loss rate is higher and the wind is slower, 
the projected size of the emission region increases leading to
larger H$\alpha$ emission flux.  The same process probably
causes the $V$-band variations, but the fractional radius variations
must be smaller at the continuum forming radius because we found
in last section that 
$\triangle \ln f [V] / \triangle \ln f [{{\rm H}\alpha}] = 0.16$ so that 
$\triangle \ln r_{\tau V} = 0.16 \triangle \ln r_{\tau {\rm H}\alpha}$, 
i.e., the continuum size variations are only $16\%$ as large as 
those in H$\alpha$. It is possible that the changes in the mass-loss rate are caused by a change in the luminosity of the star which would propagate through the wind, and be observed in both the $V$-band brightness and the emission line flux of H$\alpha$.

Finally we return to the relationship of the DACs to the
short SD-phase variations.  We found a trend with
the DAC strength and the H$\alpha$ emission equivalent width.
We show in Figure~11 the temporal behavior of DAC quotient equivalent width
$W_\lambda$(DAC) (Table 1, column 10) along with scaled, running averages of the
H$\alpha$ equivalent width and $V$-band flux (Fig.~3).   
We rescaled the amplitude of the H$\alpha$ flux by 
$\triangle W_{\lambda} {\rm (DAC)} = \triangle W_{\lambda}{\rm (corr)} / 20.46$ 
and the photometric light curve was rescaled by 
$\triangle W_{\lambda} {\rm (DAC)} = \triangle f_{V}\times 22.56$. 
These curves were then shifted vertically to match the $W_{\lambda}({\rm DAC})$ points. 
We see that the DAC strength variations track both the H$\alpha$ and $V$-band flux
variations, suggesting that the DACs are related in some way to the
short SD-phase changes.  For example, we see that some of the strongest DACs 
were observed when the H$\alpha$ emission was strong (around HJD 2,452,500; MJD 5.25$\cdot 10^4$ in our plots) 
and the DACs were very faint or absent when the emission was weak (around HJD~2,450,500).

\begin{figure}
\begin{center}
\includegraphics[angle=90,height = 12cm]{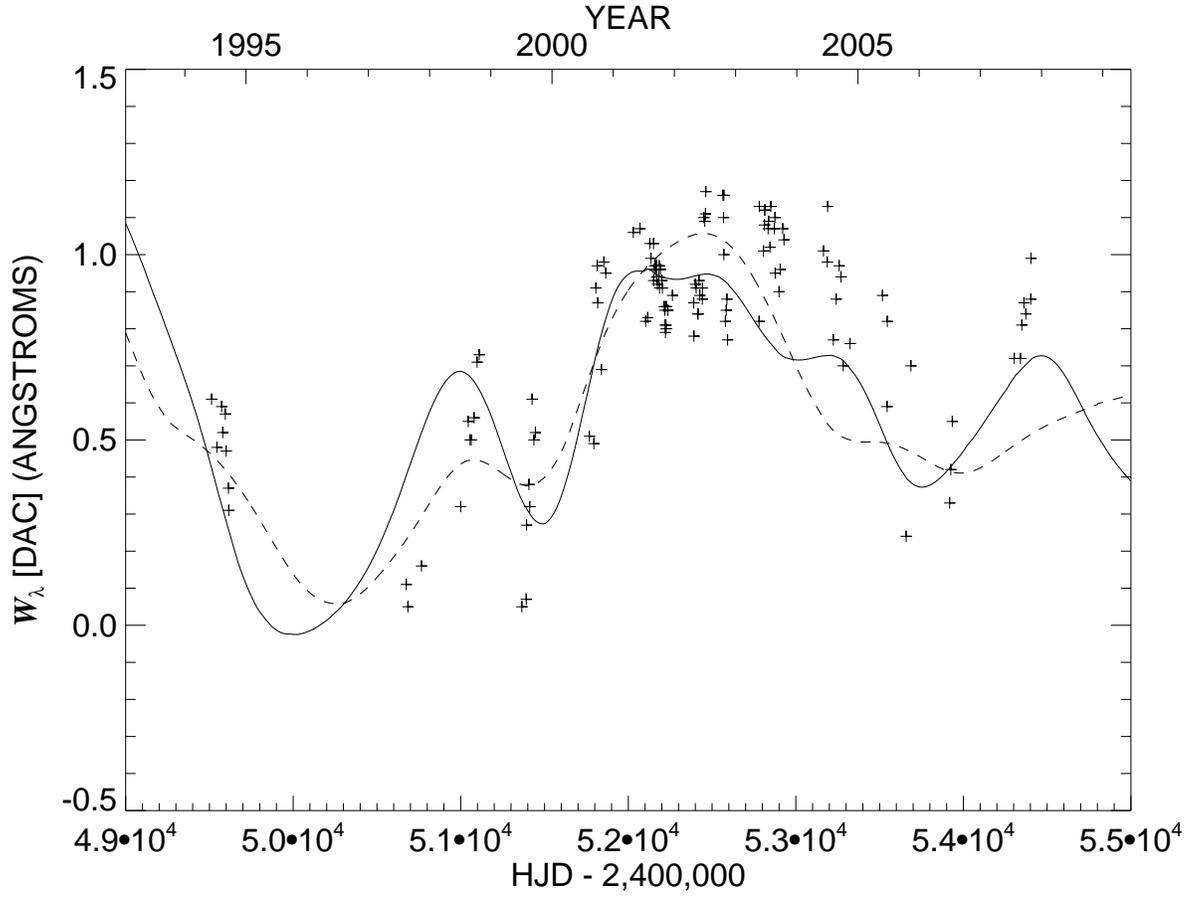}
\end{center}
\caption{The temporal variations of the relative equivalent width of the DACs. We overplot the running average of the $V$-band photometry (solid line) and the H$\alpha$ corrected equivalent width (dashed line) that were both rescaled and shifted to match the DAC variability (see text). The DAC strength appears to follow the variations associated with the short-SD phase.}
\label{fig11}
\end{figure}

\placefigure{fig11}

The DAC phenomenon is primarily observed in the UV wind lines of hot stars 
(Kaper et al.\ 1999; Puls et al.\ 2008), and, in fact, DACs are only rarely seen
in the H$\alpha$ profile where wind-related variations are usually due to changes in the dense and 
slower moving wind close to the star (Kaper et al.\ 1997; Markova et al.\ 2005). The only comparable DAC observed in H$\alpha$ was in HD 92207 (Kaufer et al.\ 1996) with a lifetime of $\sim 150$ d and only one DAC observed, so the recurrence time is unknown. This star is cooler than P Cygni and is a ``normal" supergiant, compared to the luminous blue variable nature of P Cyg.
The DACs observed in the UV wind lines of O-stars are first seen at 
velocities of 0.2 to 0.4 $v_\infty$ and then they migrate blueward to 
$v_\infty$ on time scales of a day or so, exhibiting accelerations that
are much slower than expected for the wind velocity law (Kaper et al.\ 1999).
Many of these same features are seen in the DACs in H$\alpha$ for P~Cygni, 
although they occur on vastly longer time scales.  For example, the 
wind flow time scales as $R_\star / v_\infty$, and while the wind gas will 
accelerate from $0.1 v_\infty$ to $0.9 v_\infty$ in 0.5~d for an O-star 
like $\xi$~Per (Kaper et al.\ 1999), it will take some 43~d for the wind
of P~Cygni.  However, this flow time is very short compared to the 
longevity of the DACs ($10^3$~d), so we are led to the same conclusion 
found for the O-stars, namely that the DACs represent some kind of 
perturbation in the wind through which the gas flows.  Changes in wind velocity 
and/or mass loss rate can cause shocks and create structure in the wind,
and these structures produce density enhancements and/or velocity plateaus
that imprint DACs in the wind lines (Fullerton \& Owocki 1992; 
Runacres \& Owocki 2002; Puls et al.\ 2008).  

Current theory suggests 
that the DACs are related to large scale spiral features in the wind
known as co-rotating interaction regions that are formed at the intersection 
of fast and slower outflows, which develop from some inhomogeneity in 
the stellar photosphere (for example, pulsation or spots; 
Cranmer \& Owocki 1996; Lobel \& Blomme 2008).  In these models, it
is the slow transit of these equatorially centered regions across the photosphere
that creates the DACs in the absorption cores but has little influence 
on the emission parts of the wind line (Dessart 2004; Lobel \& Blomme 2008).
However, in the case of P~Cygni, we find that emission parts do appear 
to strengthen when DACs are prominent (Fig.~11), and this indicates that
the wind perturbation profoundly affects both wind gas surrounding the 
star and the wind gas projected against the photosphere.  Thus, we suggest that 
the structures causing the DACs in P~Cygni may be more spherically symmetric 
than assumed in the geometry of the co-rotating interaction regions. 
In some models the seed perturbation occurs at a fixed longitude on the 
star, so that a new wind structure appears each time the star rotates 
(although for the best studied case of HD~64760, Lobel \& Blomme 2008
argue that the originating spots must rotate some five times slower than
the star in order to fit models to the observations).  If this is the 
case for P~Cygni, then the 1700~d DAC recurrence time may be related to the 
star's rotational period.  Markova (2000) found an upper limit of 
100~d for the rotational period based upon estimates of the stellar radius and 
the projected rotational velocity.  However, the line broadening in early 
supergiants may be dominated by turbulence rather than rotation 
(Howarth et al.~1997, 2007; Markova \& Puls 2008), so a longer rotational 
period remains a possibility.  However, regardless of the origin of 
the DACs, their close relation with the emission line and continuum flux 
variations (Fig.~11) suggests that much of the short-SD variability 
is caused by propagating structural perturbations in the outer atmosphere
of the star.  

The discovery of the short SD-phase variations in P Cygni and its
relationship to the wind velocity and DAC occurrence in the wind is a new
observational result that warrants future investigation both for this star
and other LBVs.  The changing characteristics over these long timescales may
eventually lead to a better understanding of the LBV stage of evolution and the
underlying physics of their winds and circumstellar environments.  
We are currently pursuing a three year, spectroscopic monitoring program of 
Galactic and Magellanic Cloud LBVs. Such long-term observations will reveal if 
DACs are present in all LBVs and will show whether the variations found in P~Cygni 
are a general phenomena among LBVs.

\acknowledgments 
We are grateful to an anonymous referee for important and helpful comments 
on an earlier version of this paper that greatly improved the quality of the paper. 
We acknowledge with thanks the variable star observations from the AAVSO International Database contributed by observers worldwide and used in this research. Support for Ritter Astrophysical Research Center during the time of the observations was provided by the Fund for Astrophysical Research, the American Astronomical Society Small Grants Program as well as the National Science Foundation Program for Research and Education with Small Telescopes (NSF-PREST) under grant AST-0440784 (NM). This work was also supported by the National Science Foundation under grants AST-0606861 and AST-1009080 (DG). Institutional support has been provided from the GSU College of Arts and Sciences and from the Research Program Enhancement fund of the Board of Regents of the University System of Georgia, administered through the GSU Office of the Vice President for Research. N. Markova acknowledges the Bulgarian NFSR grant DO 02-85. We are grateful for all this support. 

{\it Facilities:} \facility{Ritter Observatory} \facility{AAVSO}

%%%%%%%%%%%%%%%%%%%%%%%%%%%%%%%%%%%%%%%%%%%%%%%%%%%%%%%%%%%%%% 

%%%%%%%%%%%%%%%%%%%%%%%%%%%%%%%%%%%%%%%%%%%%%%%%%%%%%%%%%%%%%%% 
% Tables 

\begin{deluxetable}{cccccccccc}
\tablecaption{H$\alpha$ Measurements}
\tablewidth{0pt}
%\tableheight{0pt}
\rotate
\centering
\tablehead{
  \colhead{HJD}         &
  \colhead{$W_{\lambda}$}  &
  \colhead{}  &  
  \colhead{$W_{\lambda}$}  &
  \colhead{} &
  \colhead{} &
  \colhead{} &
  \colhead{$V_r$} &
  \colhead{$V_r$} &
  \colhead{$W_{\lambda}$} \\  

  \colhead{--2,400,000}         &
  \colhead{(net)}  &  
  \colhead{$V$}         &  
  \colhead{(corr)}         &  
  \colhead{$I_p$}         &  
  \colhead{FWHM}         &
  \colhead{$\triangle V_r$}  & 
  \colhead{(min)}    & 
  \colhead{(DAC)}  &
  \colhead{(DAC)} \\ 

  \colhead{(d)}  &
  \colhead{(\AA )}  &
  \colhead{(mag)} &
  \colhead{(\AA )} &
  \colhead{(norm.)} &
  \colhead{(km s$^{-1}$)} &
  \colhead{(km s$^{-1}$)} &
  \colhead{(km s$^{-1}$)} &
  \colhead{(km s$^{-1}$)} &
  \colhead{(\AA )} \\
 
  \colhead{(1)}  &
  \colhead{(2)}  &
  \colhead{(3)} &
  \colhead{(4)} &
  \colhead{(5)} &
  \colhead{(6)} &
  \colhead{(7)} &
  \colhead{(8)} &
  \colhead{(9)} &
  \colhead{(10)}}    

\startdata
49512.730	&	89.8	&	4.79	&	90.6	&	21.1	&	169.8	&	\phs0.3	&	 $-$203.3	&	 $-$151.1	&	0.61	\\
49544.770	&	91.5	&	4.81	&	90.7	&	20.7	&	177.4	&	 $-$2.5	&	 $-$203.4	&	 $-$159.6	&	0.48	\\
49572.755	&	76.3	&	4.72	&	82.1	&	18.3	&	183.6	&	 $-$0.8	&	 $-$203.7	&	 $-$156.2	&	0.59	\\
49579.717	&	78.1	&	4.74	&	82.5	&	17.9	&	177.4	&	 $-$0.5	&	 $-$204.7	&	 $-$154.4	&	0.52	\\
49594.692	&	80.2	&	4.81	&	79.5	&	17.3	&	182.0	&	 $-$1.0	&	 $-$205.0	&	 $-$156.9	&	0.57	\\
49599.701	&	83.1	&	4.83	&	80.8	&	17.2	&	179.0	&	 $-$1.5	&	 $-$205.6	&	 $-$157.7	&	0.47	\\
49613.687	&	83.0	&	4.86	&	78.5	&	17.9	&	175.9	&	 $-$4.3	&	 $-$206.5	&	 $-$162.2	&	0.37	\\
49615.743	&	84.5	&	4.86	&	80.0	&	18.1	&	179.0	&	 $-$5.0	&	 $-$206.2	&	 $-$167.3	&	0.31	\\
49628.561	&	83.4	&	4.85	&	79.6	&	17.5	&	186.6	&	 $-$5.9	&	 $-$206.4	&	\nodata	&	\nodata	\\
49629.640	&	79.7	&	4.85	&	76.1	&	16.9	&	188.2	&	 $-$6.1	&	 $-$206.7	&	\nodata	&	\nodata	\\
49638.584	&	80.1	&	4.82	&	78.6	&	16.7	&	192.7	&	 $-$7.5	&	 $-$206.5	&	\nodata	&	\nodata	\\
49664.618	&	75.1	&	4.83	&	73.1	&	14.6	&	200.4	&	 $-$3.7	&	 $-$206.5	&	\nodata	&	\nodata	\\
50003.651	&	67.2	&	4.78	&	68.4	&	14.0	&	189.7	&	 $-$7.4	&	 $-$209.7	&	\nodata	&	\nodata	\\
50013.574	&	70.8	&	4.77	&	72.8	&	14.5	&	192.7	&	 $-$7.9	&	 $-$210.8	&	\nodata	&	\nodata	\\
50045.460	&	77.5	&	4.84	&	74.7	&	15.2	&	191.2	&	 $-$4.4	&	 $-$210.7	&	\nodata	&	\nodata	\\
50074.495	&	72.2	&	\nodata	&	\nodata	&	\nodata	&	195.8	&	 $-$4.9	&	 $-$211.7	&	\nodata	&	\nodata	\\
50268.843	&	66.4	&	4.75	&	69.5	&	14.0	&	186.6	&	 $-$6.6	&	 $-$215.3	&	\nodata	&	\nodata	\\
50321.633	&	69.3	&	4.71	&	75.3	&	15.7	&	174.4	&	 $-$1.4	&	 $-$212.3	&	\nodata	&	\nodata	\\
50604.817	&	77.4	&	4.82	&	76.0	&	15.5	&	182.0	&	 $-$0.5	&	 $-$212.6	&	\nodata	&	\nodata	\\
50609.794	&	75.5	&	4.82	&	74.1	&	15.8	&	174.4	&	 $-$1.0	&	 $-$213.0	&	\nodata	&	\nodata	\\
50614.794	&	79.5	&	4.82	&	78.0	&	16.7	&	179.0	&	 $-$0.8	&	 $-$212.0	&	\nodata	&	\nodata	\\
50626.794	&	81.2	&	4.82	&	79.7	&	17.4	&	180.5	&	 $-$2.2	&	 $-$212.5	&	\nodata	&	\nodata	\\
50631.737	&	73.7	&	4.82	&	72.4	&	15.8	&	182.0	&	 $-$2.5	&	 $-$212.8	&	\nodata	&	\nodata	\\
50649.843	&	78.4	&	\nodata	&	\nodata	&	\nodata	&	188.2	&	 $-$2.7	&	 $-$211.6	&	\nodata	&	\nodata	\\
50674.708	&	71.5	&	4.78	&	72.8	&	15.4	&	192.7	&	 $-$1.5	&	 $-$210.9	&	 $-$126.4	&	0.11	\\
50684.641	&	70.7	&	4.78	&	72.0	&	14.5	&	189.7	&	 $-$3.6	&	 $-$211.1	&	 $-$128.1	&	0.05	\\
50717.689	&	68.2	&	4.75	&	71.4	&	16.6	&	174.4	&	 $-$4.5	&	 $-$211.4	&	\nodata	&	\nodata	\\
50765.545	&	75.0	&	\nodata	&	\nodata	&	\nodata	&	185.1	&	 $-$0.4	&	 $-$211.0	&	 $-$142.3	&	0.16	\\
50999.782	&	81.0	&	4.70	&	88.8	&	20.9	&	160.6	&	\phs0.1	&	 $-$210.7	&	 $-$146.5	&	0.32	\\
51044.769	&	82.4	&	4.76	&	85.5	&	19.2	&	163.7	&	 $-$2.7	&	 $-$211.1	&	 $-$144.4	&	0.55	\\
51055.623	&	81.2	&	4.74	&	85.8	&	19.6	&	165.2	&	 $-$2.4	&	 $-$209.8	&	 $-$145.9	&	0.50	\\
51062.680	&	74.7	&	4.72	&	80.4	&	18.4	&	165.2	&	 $-$1.9	&	 $-$209.0	&	 $-$148.5	&	0.50	\\
51079.602	&	77.8	&	4.72	&	83.7	&	18.3	&	166.7	&	\phs0.8	&	 $-$208.0	&	 $-$145.9	&	0.56	\\
51097.522	&	79.2	&	4.75	&	82.9	&	18.3	&	165.2	&	\phs2.1	&	 $-$208.3	&	 $-$142.8	&	0.71	\\
51110.586	&	82.0	&	4.76	&	85.1	&	19.3	&	163.7	&	\phs0.5	&	 $-$207.9	&	 $-$142.0	&	0.73	\\
51336.757	&	83.4	&	4.85	&	79.6	&	16.5	&	192.7	&	 $-$5.3	&	 $-$206.5	&	\nodata	&	\nodata	\\
51338.791	&	86.2	&	4.87	&	80.8	&	16.6	&	194.3	&	 $-$5.8	&	 $-$207.8	&	\nodata	&	\nodata	\\
51348.808	&	85.8	&	4.89	&	79.0	&	16.3	&	194.3	&	 $-$3.1	&	 $-$207.0	&	\nodata	&	\nodata	\\
51364.768	&	76.1	&	4.83	&	74.0	&	15.1	&	194.3	&	 $-$1.6	&	 $-$207.0	&	 $-$162.9	&	0.05	\\
51390.752	&	72.2	&	4.82	&	70.9	&	14.5	&	192.7	&	 $-$1.2	&	 $-$205.6	&	 $-$163.2	&	0.07	\\
51392.778	&	74.7	&	4.82	&	73.3	&	14.5	&	195.8	&	 $-$1.6	&	 $-$207.0	&	 $-$163.1	&	0.27	\\
51406.744	&	69.7	&	4.74	&	73.7	&	14.4	&	192.7	&	\phs0.0	&	 $-$204.8	&	 $-$157.3	&	0.38	\\
51411.784	&	70.0	&	4.72	&	75.4	&	15.5	&	185.1	&	 $-$2.6	&	 $-$204.7	&	 $-$162.9	&	0.32	\\
51425.618	&	70.0	&	4.74	&	74.0	&	14.8	&	185.1	&	 $-$2.6	&	 $-$203.7	&	 $-$163.0	&	0.61	\\
51436.675	&	73.8	&	4.78	&	75.2	&	15.2	&	180.5	&	 $-$1.2	&	 $-$203.4	&	 $-$162.0	&	0.50	\\
51445.630	&	76.3	&	4.82	&	74.9	&	15.2	&	185.1	&	\phs0.3	&	 $-$203.7	&	 $-$160.4	&	0.52	\\
51767.661	&	84.0	&	4.78	&	85.6	&	19.9	&	154.5	&	\phs3.9	&	 $-$207.5	&	 $-$141.7	&	0.51	\\
51794.670	&	79.5	&	4.74	&	84.0	&	19.4	&	156.0	&	\phs3.3	&	 $-$206.5	&	 $-$142.8	&	0.49	\\
51806.567	&	80.8	&	4.76	&	83.8	&	19.4	&	156.0	&	\phs1.2	&	 $-$208.5	&	 $-$139.9	&	0.91	\\
51814.620	&	68.2	&	4.76	&	70.8	&	17.4	&	160.6	&	\phs1.7	&	 $-$209.7	&	 $-$123.5	&	0.97	\\
51817.658	&	82.5	&	4.75	&	86.4	&	19.4	&	156.0	&	\phs2.0	&	 $-$207.0	&	 $-$142.0	&	0.87	\\
51839.556	&	77.6	&	4.71	&	84.3	&	20.2	&	162.1	&	\phs0.5	&	 $-$207.5	&	 $-$141.5	&	0.69	\\
51854.489	&	78.5	&	4.73	&	83.7	&	19.1	&	159.1	&	\phs0.1	&	 $-$205.8	&	 $-$144.2	&	0.98	\\
51865.490	&	72.2	&	4.71	&	78.4	&	18.2	&	154.5	&	\phs0.1	&	 $-$205.6	&	 $-$144.8	&	0.95	\\
52029.826	&	91.1	&	4.77	&	93.7	&	22.8	&	143.8	&	\phs6.5	&	 $-$206.2	&	 $-$137.2	&	1.06	\\
52069.810	&	95.2	&	4.75	&	99.7	&	23.9	&	149.9	&	\phs6.0	&	 $-$203.3	&	 $-$139.8	&	1.07	\\
52103.753	&	87.8	&	4.68	&	98.1	&	22.7	&	157.6	&	\phs3.4	&	 $-$200.0	&	 $-$145.1	&	0.82	\\
52115.730	&	90.3	&	4.72	&	97.2	&	21.9	&	156.0	&	\phs1.9	&	 $-$199.7	&	 $-$145.7	&	0.83	\\
52128.677	&	85.7	&	4.70	&	94.0	&	21.7	&	151.4	&	\phs2.6	&	 $-$199.5	&	 $-$145.4	&	1.03	\\
52135.638	&	83.7	&	4.71	&	90.9	&	21.1	&	146.9	&	\phs2.4	&	 $-$200.8	&	 $-$142.3	&	0.99	\\
52150.688	&	86.9	&	4.74	&	91.8	&	21.6	&	140.7	&	\phs2.6	&	 $-$200.8	&	 $-$143.2	&	1.03	\\
52151.618	&	86.8	&	4.74	&	91.7	&	21.1	&	145.3	&	\phs3.5	&	 $-$200.6	&	 $-$142.6	&	0.93	\\
52154.542	&	85.9	&	4.74	&	90.8	&	21.1	&	145.3	&	\phs3.0	&	 $-$201.1	&	 $-$143.2	&	0.96	\\
52158.587	&	86.3	&	4.74	&	91.2	&	21.3	&	143.8	&	\phs3.2	&	 $-$201.0	&	 $-$142.6	&	0.97	\\
52163.578	&	85.7	&	4.75	&	89.7	&	21.1	&	142.3	&	\phs3.6	&	 $-$201.1	&	 $-$141.4	&	0.97	\\
52165.550	&	87.4	&	4.75	&	91.5	&	21.4	&	142.3	&	\phs4.3	&	 $-$200.9	&	 $-$142.3	&	0.96	\\
52168.595	&	86.7	&	4.78	&	88.3	&	20.6	&	137.7	&	\phs4.6	&	 $-$200.7	&	 $-$139.8	&	0.94	\\
52169.558	&	87.7	&	4.79	&	88.5	&	20.8	&	137.7	&	\phs4.0	&	 $-$201.7	&	 $-$141.6	&	0.97	\\
52173.616	&	88.0	&	4.79	&	88.8	&	21.0	&	139.2	&	\phs4.7	&	 $-$200.9	&	 $-$141.2	&	0.92	\\
52182.549	&	91.2	&	4.80	&	91.2	&	21.7	&	136.1	&	\phs4.2	&	 $-$201.1	&	 $-$140.1	&	0.94	\\
52184.554	&	90.3	&	4.80	&	90.3	&	21.7	&	136.1	&	\phs5.1	&	 $-$199.9	&	 $-$140.9	&	0.97	\\
52186.594	&	92.2	&	4.80	&	92.2	&	22.3	&	136.1	&	\phs4.9	&	 $-$200.1	&	 $-$140.2	&	0.91	\\
52192.594	&	94.6	&	4.79	&	95.5	&	22.5	&	136.1	&	\phs4.2	&	 $-$200.8	&	 $-$140.5	&	0.96	\\
52200.512	&	94.9	&	4.79	&	95.8	&	23.4	&	137.7	&	\phs4.2	&	 $-$200.9	&	 $-$138.4	&	0.93	\\
52203.478	&	96.6	&	4.79	&	97.5	&	23.3	&	140.7	&	\phs3.8	&	 $-$201.1	&	 $-$138.7	&	0.91	\\
52214.476	&	95.8	&	4.80	&	95.8	&	23.5	&	140.7	&	\phs3.1	&	 $-$201.4	&	 $-$136.9	&	0.86	\\
52217.596	&	97.2	&	4.80	&	97.2	&	23.4	&	146.9	&	\phs2.6	&	 $-$200.7	&	 $-$137.3	&	0.85	\\
52220.481	&	94.5	&	4.80	&	94.5	&	23.2	&	145.3	&	\phs2.9	&	 $-$200.5	&	 $-$137.3	&	0.81	\\
52221.510	&	96.8	&	4.79	&	97.7	&	24.1	&	143.8	&	\phs2.5	&	 $-$200.6	&	 $-$136.9	&	0.79	\\
52224.476	&	96.1	&	4.79	&	97.0	&	23.9	&	148.4	&	\phs3.8	&	 $-$198.7	&	 $-$138.1	&	0.81	\\
52225.464	&	95.8	&	4.79	&	96.7	&	24.3	&	145.3	&	\phs3.4	&	 $-$198.7	&	 $-$138.1	&	0.80	\\
52226.462	&	93.8	&	4.79	&	94.7	&	23.3	&	146.9	&	\phs3.4	&	 $-$198.8	&	 $-$139.5	&	0.86	\\
52236.465	&	94.5	&	4.78	&	96.3	&	23.7	&	148.4	&	\phs3.5	&	 $-$198.1	&	 $-$139.6	&	0.85	\\
52263.477	&	98.0	&	\nodata	&	\nodata	&	\nodata	&	159.1	&	\phs3.3	&	 $-$198.0	&	 $-$139.4	&	0.89	\\
52390.821	&	89.2	&	\nodata	&	\nodata	&	\nodata	&	145.3	&	\phs4.5	&	 $-$196.6	&	 $-$141.9	&	0.87	\\
52391.857	&	87.7	&	\nodata	&	\nodata	&	\nodata	&	151.4	&	\phs4.5	&	 $-$197.0	&	 $-$140.5	&	0.78	\\
52399.803	&	90.8	&	\nodata	&	\nodata	&	\nodata	&	154.5	&	\phs3.9	&	 $-$196.5	&	 $-$138.7	&	0.92	\\
52404.875	&	89.2	&	\nodata	&	\nodata	&	\nodata	&	154.5	&	\phs3.2	&	 $-$198.0	&	 $-$138.2	&	0.91	\\
52413.831	&	92.5	&	\nodata	&	\nodata	&	\nodata	&	157.6	&	\phs2.4	&	 $-$196.4	&	 $-$141.4	&	0.84	\\
52416.779	&	90.6	&	\nodata	&	\nodata	&	\nodata	&	159.1	&	\phs1.4	&	 $-$197.3	&	 $-$140.2	&	0.84	\\
52421.800	&	94.9	&	\nodata	&	\nodata	&	\nodata	&	163.7	&	\phs0.7	&	 $-$196.8	&	 $-$140.4	&	0.93	\\
52426.786	&	92.0	&	\nodata	&	\nodata	&	\nodata	&	163.7	&	\phs1.1	&	 $-$196.2	&	 $-$140.8	&	0.89	\\
52441.792	&	83.8	&	\nodata	&	\nodata	&	\nodata	&	165.2	&	\phs1.6	&	 $-$194.6	&	 $-$145.1	&	0.88	\\
52442.776	&	81.3	&	\nodata	&	\nodata	&	\nodata	&	162.1	&	\phs2.3	&	 $-$192.2	&	 $-$143.7	&	0.91	\\
52450.756	&	81.9	&	\nodata	&	\nodata	&	\nodata	&	163.7	&	\phs1.6	&	 $-$194.6	&	 $-$145.2	&	1.10	\\
52456.691	&	79.7	&	\nodata	&	\nodata	&	\nodata	&	159.1	&	\phs1.8	&	 $-$193.2	&	 $-$145.8	&	1.09	\\
52460.804	&	83.3	&	4.71	&	90.5	&	19.7	&	165.2	&	\phs2.0	&	 $-$192.2	&	 $-$146.1	&	1.11	\\
52462.751	&	83.1	&	4.71	&	90.3	&	19.6	&	163.7	&	\phs1.0	&	 $-$193.9	&	 $-$146.1	&	1.17	\\
52564.557	&	95.7	&	4.72	&     103.0\phn &	23.8	&	163.7	&	\phs8.0	&	 $-$181.4	&	 $-$146.4	&	1.16	\\
52568.508	&	92.9	&	4.74	&	98.2	&	22.9	&	159.1	&	 $-$3.3	&	 $-$192.9	&	 $-$137.8	&	1.10	\\
52570.487	&	93.9	&	4.75	&	98.3	&	23.1	&	160.6	&	 $-$1.0	&	 $-$191.2	&	 $-$148.0	&	1.00	\\
52570.522	&	95.8	&	4.75	&     100.3\phn &	23.4	&	157.6	&	 $-$1.4	&	 $-$191.4	&	 $-$148.1	&	1.16	\\
52579.490	&	91.8	&	4.78	&	93.5	&	22.0	&	163.7	&	 $-$1.3	&	 $-$191.4	&	 $-$147.9	&	0.82	\\
52584.499	&	95.4	&	4.81	&	94.5	&	22.1	&	163.7	&	\phs0.7	&	 $-$190.0	&	 $-$147.1	&	0.85	\\
52588.491	&	95.7	&	4.81	&	94.8	&	22.6	&	163.7	&	\phs0.4	&	 $-$190.0	&	 $-$146.2	&	0.88	\\
52592.504	&	94.2	&	4.81	&	93.3	&	22.0	&	165.2	&	 $-$1.4	&	 $-$191.5	&	 $-$138.4	&	0.77	\\
52780.814	&	82.9	&	\nodata	&	\nodata	&	\nodata	&	157.6	&	\phs2.2	&	 $-$195.0	&	 $-$146.0	&	0.82	\\
52781.812	&	81.3	&	\nodata	&	\nodata	&	\nodata	&	153.0	&	\phs1.9	&	 $-$194.3	&	 $-$146.2	&	1.13	\\
52806.774	&	81.9	&	4.80	&	81.9	&	18.7	&	159.1	&	\phs0.0	&	 $-$195.7	&	 $-$146.0	&	1.01	\\
52812.789	&	86.1	&	4.78	&	87.7	&	19.9	&	159.1	&	 $-$0.7	&	 $-$196.7	&	 $-$146.1	&	1.08	\\
52813.748	&	87.6	&	4.78	&	89.2	&	20.1	&	160.6	&	 $-$3.4	&	 $-$201.0	&	 $-$144.1	&	1.12	\\
52815.647	&	88.4	&	4.78	&	90.0	&	20.6	&	159.1	&	\phs1.1	&	 $-$195.5	&	 $-$146.4	&	1.12	\\
52834.754	&	85.5	&	4.78	&	87.1	&	19.7	&	163.7	&	\phs4.2	&	 $-$193.5	&	 $-$147.9	&	1.07	\\
52837.646	&	83.0	&	4.78	&	84.5	&	19.1	&	163.7	&	 $-$0.4	&	 $-$196.8	&	 $-$146.8	&	1.09	\\
52845.679	&	86.8	&	4.77	&	89.2	&	20.1	&	166.7	&	 $-$5.9	&	 $-$204.6	&	 $-$145.5	&	1.02	\\
52850.628	&	84.3	&	4.77	&	86.7	&	19.6	&	163.7	&	\phs2.2	&	 $-$194.8	&	 $-$149.1	&	1.13	\\
52871.608	&	87.8	&	4.78	&	89.4	&	20.3	&	162.1	&	 $-$1.0	&	 $-$197.2	&	 $-$148.7	&	1.07	\\
52875.607	&	87.5	&	4.77	&	90.0	&	20.4	&	163.7	&	 $-$0.9	&	 $-$197.3	&	 $-$149.3	&	1.10	\\
52876.683	&	87.2	&	4.77	&	89.6	&	20.6	&	162.1	&	 $-$2.3	&	 $-$197.2	&	 $-$148.5	&	0.95	\\
52899.677	&	81.9	&	4.77	&	84.2	&	18.6	&	171.3	&	 $-$3.5	&	 $-$195.7	&	 $-$149.2	&	0.90	\\
52906.649	&	82.9	&	4.79	&	83.7	&	18.6	&	171.3	&	 $-$5.2	&	 $-$196.8	&	 $-$150.4	&	0.96	\\
52921.570	&	82.2	&	4.77	&	84.5	&	18.9	&	169.8	&	 $-$0.8	&	 $-$192.9	&	 $-$151.0	&	1.07	\\
52929.570	&	79.2	&	4.77	&	81.4	&	17.9	&	169.8	&	\phs3.5	&	 $-$187.8	&	 $-$152.6	&	1.04	\\
53164.826	&	68.6	&	4.77	&	70.5	&	15.0	&	169.8	&  $-$10.4\phn	&	 $-$208.9	&	 $-$142.6	&	1.01	\\
53187.777	&	79.0	&	4.78	&	80.5	&	16.5	&	175.9	&	 $-$7.3	&	 $-$204.8	&	 $-$158.9	&	0.98	\\
53189.727	&	77.9	&	4.77	&	80.1	&	16.6	&	172.9	&	 $-$4.6	&	 $-$206.5	&	 $-$156.5	&	1.13	\\
53223.686	&	76.3	&	4.72	&	82.1	&	16.9	&	180.5	&	 $-$3.6	&	 $-$205.1	&	 $-$156.2	&	0.77	\\
53240.676	&	78.8	&	4.78	&	80.3	&	16.8	&	177.4	&	 $-$8.3	&	 $-$207.7	&	 $-$162.2	&	0.88	\\
53258.635	&	75.1	&	4.77	&	77.2	&	15.9	&	175.9	&	 $-$4.2	&	 $-$206.0	&	 $-$162.6	&	0.97	\\
53269.540	&	77.1	&	4.74	&	81.5	&	16.6	&	179.0	&	 $-$3.1	&	 $-$204.8	&	 $-$160.2	&	0.94	\\
53282.531	&	73.1	&	4.73	&	78.0	&	16.2	&	171.3	&	 $-$3.9	&	 $-$205.7	&	 $-$157.6	&	0.70	\\
53323.501	&	74.9	&	4.74	&	79.2	&	17.4	&	165.2	&	 $-$9.3	&	 $-$209.9	&	 $-$154.7	&	0.76	\\
53499.871	&	78.8	&	4.74	&	83.3	&	18.4	&	165.2	&	 $-$5.4	&	 $-$206.7	&	\nodata	&	\nodata	\\
53516.855	&	81.0	&	4.74	&	85.6	&	18.2	&	168.3	&  $-$10.1\phn	&	 $-$210.3	&	 $-$151.1	&	0.89	\\
53544.699	&	81.9	&	4.75	&	85.8	&	19.1	&	166.7	&	 $-$2.9	&	 $-$203.8	&	 $-$154.2	&	0.59	\\
53545.746	&	79.7	&	4.75	&	83.5	&	19.0	&	159.1	&	\phs1.8	&	 $-$202.5	&	 $-$153.8	&	0.82	\\
53559.761	&	80.0	&	4.73	&	85.3	&	19.5	&	157.6	&	 $-$2.6	&	 $-$207.8	&	\nodata	&	\nodata	\\
53623.728	&	85.0	&	4.81	&	84.2	&	18.8	&	174.4	&	 $-$1.0	&	 $-$203.3	&	\nodata	&	\nodata	\\
53644.603	&	82.7	&	4.84	&	79.7	&	17.2	&	186.6	&	 $-$6.8	&	 $-$210.1	&	 $-$155.2	&	 $-$0.18	\\
53658.533	&	79.8	&	4.80	&	79.8	&	16.8	&	188.2	&	 $-$3.3	&	 $-$205.9	&	 $-$161.1	&	0.24	\\
53686.551	&	72.1	&	4.75	&	75.5	&	15.5	&	192.7	&	 $-$6.2	&	 $-$206.6	&	 $-$157.9	&	0.70	\\
53892.814	&	77.1	&	4.82	&	75.7	&	16.2	&	165.2	&	 $-$4.2	&	 $-$208.5	&	\nodata	&	\nodata	\\
53902.819	&	80.2	&	4.82	&	78.7	&	17.0	&	169.8	&	 $-$2.8	&	 $-$207.6	&	\nodata	&	\nodata	\\
53912.775	&	76.3	&	4.77	&	78.4	&	16.2	&	174.4	&	 $-$3.7	&	 $-$207.2	&	\nodata	&	\nodata	\\
53917.754	&	79.4	&	4.74	&	83.9	&	17.4	&	179.0	&	 $-$3.5	&	 $-$206.4	&	 $-$156.0	&	0.33	\\
53924.778	&	79.5	&	4.77	&	81.7	&	17.1	&	180.5	&	\phs1.0	&	 $-$204.1	&	 $-$158.4	&	0.42	\\
53933.735	&	75.9	&	4.78	&	77.3	&	15.9	&	175.9	&	 $-$4.4	&	 $-$204.8	&	 $-$156.6	&	0.55	\\
54003.601	&	70.8	&	4.74	&	74.8	&	15.4	&	179.0	&	 $-$2.0	&	 $-$205.2	&	\nodata	&	\nodata	\\
54273.799	&	76.1	&	4.80	&	76.1	&	18.0	&	160.6	&	\phs1.2	&	 $-$203.9	&	\nodata	&	\nodata	\\
54294.753	&	78.7	&	4.79	&	79.4	&	18.5	&	172.9	&	\phs1.8	&	 $-$203.6	&	\nodata	&	\nodata	\\
54297.757	&	75.6	&	4.72	&	81.4	&	18.9	&	175.9	&	\phs0.8	&	 $-$204.1	&	\nodata	&	\nodata	\\
54303.682	&	72.8	&	4.72	&	78.4	&	18.2	&	175.9	&	\phs0.6	&	 $-$203.8	&	 $-$152.6	&	0.72	\\
54340.686	&	67.6	&	4.70	&	74.1	&	18.8	&	157.6	&	 $-$0.1	&	 $-$203.2	&	 $-$150.6	&	0.72	\\
54348.647	&	68.1	&	4.73	&	72.6	&	18.7	&	156.0	&	 $-$0.7	&	 $-$203.6	&	 $-$150.2	&	0.81	\\
54361.632	&	70.2	&	4.76	&	72.8	&	18.8	&	156.0	&	\phs0.3	&	 $-$203.0	&	 $-$149.8	&	0.87	\\
54372.584	&	64.2	&	4.74	&	67.8	&	17.5	&	156.0	&	\phs2.6	&	 $-$202.0	&	 $-$149.3	&	0.84	\\
54402.551	&	68.7	&	4.73	&	73.3	&	19.7	&	149.9	&	\phs3.5	&	 $-$202.4	&	 $-$147.5	&	0.88	\\
54403.527	&	68.4	&	4.73	&	73.0	&	19.8	&	149.9	&	\phs3.1	&	 $-$202.7	&	 $-$147.1	&	0.99	\\

\enddata
\end{deluxetable}

%%%%%%%%%%%%%%%%%%%%%%%%%%%%%%%%%%%%%%%%%%%%%%%%%%%%%%%%%%%%%%% 
% Figures 
 
\clearpage 
 
% Figure 1 $-$ H alpha vs. JD 

\clearpage 

\clearpage

\clearpage
\clearpage

\clearpage

\clearpage
 
\clearpage
\clearpage

\clearpage

\clearpage

\clearpage

%%%%%%%%%%%%%%%%%%%%%%%%%%%%%%%%%%%%%%%%%%%%%%%%%%%%%%%%%%%%%%% 

\end{document}